\documentclass[apl,reprint,superscriptaddress]{revtex4-2}
\usepackage{gensymb}
\usepackage{notes2bib}
\usepackage{graphicx}
\usepackage{textcomp}
\usepackage{verbatim}
\usepackage{amsmath}
\usepackage{setspace}
\usepackage{multirow}
\usepackage[dvipsnames]{xcolor}
\usepackage[colorlinks=true,urlcolor=blue,linkcolor=blue,citecolor=blue,a4paper]{hyperref}

\begin{document}

\title{Employing High-temperature-grown SrZrO$_3$ Buffer to Enhance the Electron Mobility in La:BaSnO$_3$-based Heterostructures}

\author{Prosper Ngabonziza}

\email[corresponding author, ]{pngabonziza@lsu.edu}
\affiliation{Department of Physics and Astronomy, Louisiana State University, Baton Rouge,  LA 70803, USA}
\affiliation{Department of Physics, University of Johannesburg, P.O. Box 524 Auckland Park 2006, Johannesburg, South Africa}
\author{Jisung Park}
\affiliation{Department of Material Science and Engineering, Cornell University, Ithaca, New York 14853, USA}
\author{Wilfried Sigle}
\affiliation{Max Planck Institute for Solid State Research, Heisenbergstr.\ 1, 70569 Stuttgart, Germany}
\author{Peter A. van Aken}
\affiliation{Max Planck Institute for Solid State Research, Heisenbergstr.\ 1, 70569 Stuttgart, Germany}
\author{Jochen Mannhart}
\affiliation{Max Planck Institute for Solid State Research, Heisenbergstr.\ 1, 70569 Stuttgart, Germany}
\author{Darrell G. Schlom}
\affiliation{Department of Material Science and Engineering, Cornell University, Ithaca, New York 14853, USA}
\affiliation{Kavli Institute at Cornell for Nanoscale Science, Ithaca, New York 14853, USA}
\affiliation{Leibniz-Institut für Kristallzüchtung, Max-Born-Str.\ 2, 12489 Berlin, Germany}

\date{\today}

\begin{abstract}
We report a synthetic route to achieve high electron mobility at room temperature in epitaxial La:BaSnO$_3$/SrZrO$_3$  heterostructures prepared on several oxide substrates. Room-temperature mobilities of 157, 145, and 143  cm$^2$V$^{-1}$s$^{-1}$ are achieved for heterostructures grown on DyScO$_3$ (110), MgO (001), and TbScO$_3$ (110) crystalline substrates, respectively. This is realized by first employing pulsed laser deposition to grow at very high temperature the SrZrO$_3$ buffer layer to reduce dislocation density in the active layer, then followed by the epitaxial growth of an overlaying La:BaSnO$_3$ active layer by molecular-beam epitaxy. Structural properties of these heterostructures are investigated, and the extracted upper limit of threading dislocations is well below $1.0\times 10^{10}$cm$^{-2}$ for buffered films on DyScO$_3$, MgO, and TbScO$_3$ substrates. The present results provide a promising route towards achieving high mobility in buffered La:BaSnO$_3$ films prepared on most, if not all,  oxide substrates with large compressive or tensile lattice mismatches to the film.
\end{abstract}

\maketitle

	The perovskite alkaline earth stannate, La-doped BaSnO$_3$ (La:BaSO$_3$), is an attractive transparent semiconductor that exhibits outstanding room-temperature  electron mobility (RT  $\mu_e$) of  $320\text{ cm}^2\text{ V}^{-1}\text{s}^{-1}$   at  a carrier density of  $ n \simeq 8.0\times 10^{19} \text{ cm}^{-3}$ in bulk single  crystals~\cite{Kim2012b,Kim2012a}. Besides its wide bandgap (3.1  eV) and unique optical properties, La:BaSO$_3$ is highly stable at high temperatures and it exhibits unique electronic properties. This makes  La:BaSO$_3$ an enticing material for the exploration of device physics in transparent high RT  $\mu_e$ field-effect transistors (FETs) and a suitable candidate material for integration in thermally stable capacitors and power electronic devices~\cite{KFujiwara_2017,Lee2017,Krishnaswamy2016,Kim2012a,Naamneh2022,SKim_2015,JYue_2018,JCheng_2021,RABucur_2012,PJisung_2020}.

\begin{figure*}[!t]
     \includegraphics[width=1\textwidth]{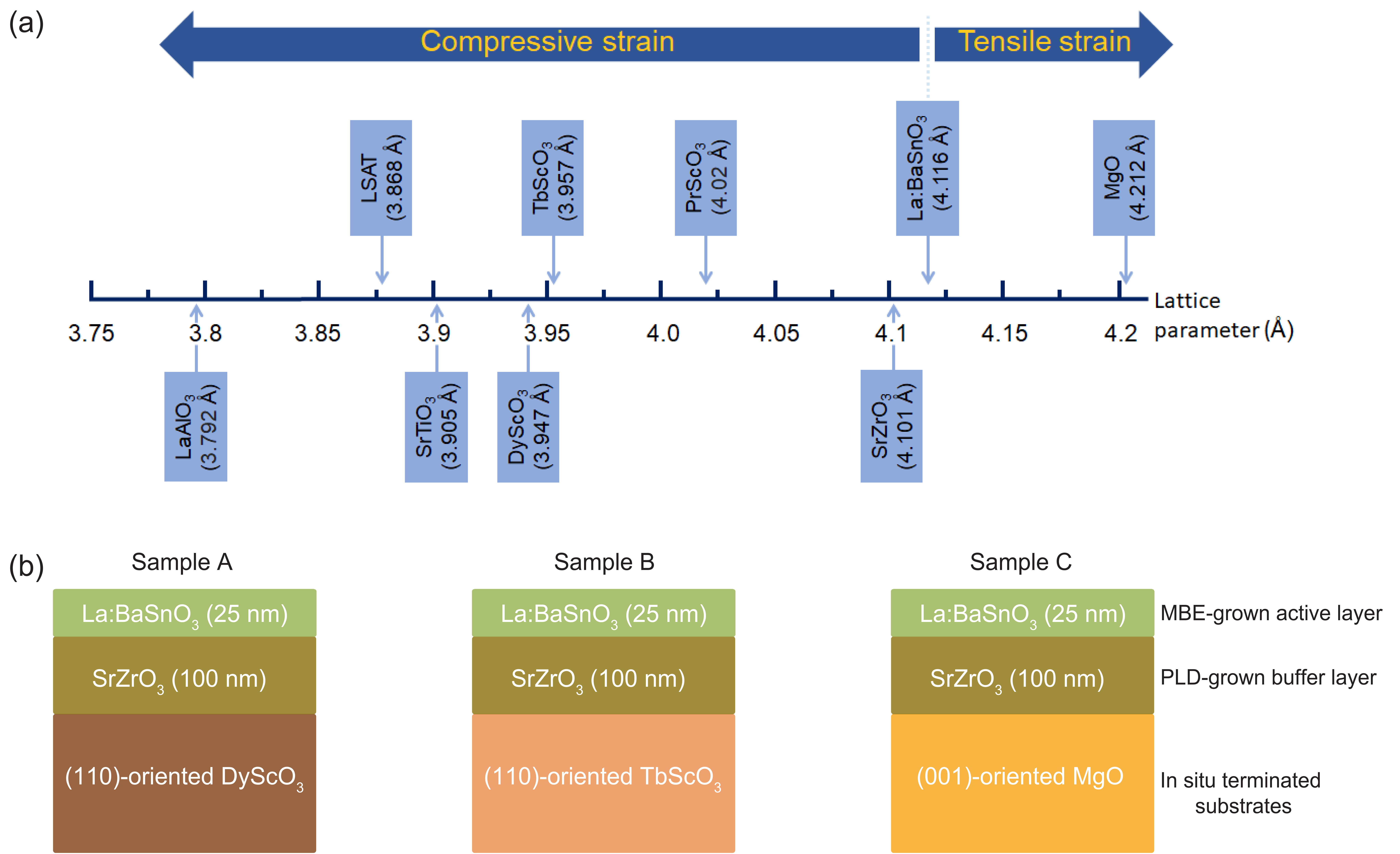}
    \caption{\small{(a) Comparative lattice constant (Å) of La:BaSnO$_3$ and  SrZrO$_3$ films with some commercially available oxide crystal substrates  within the range from $\sim 3.75$ Å to $\sim 4.22$ Å. (b) Schematic layout of the different heterostructure types investigated in this study.}} 
\label{Fig_01}
\end{figure*}

	The potential  of  La:BaSO$_3$ for oxide electronics and fundamental realization of 2-dimensional electron gases with high RT  $\mu_e$ in transparent semiconductors have triggered  considerable  interests  in  thin  films and heterostructures~\cite{NonoTchiomo2022,ZWang_2019,HCho_2019,Postiglione2021,Sanchela2018,Prakash2017c,Yu2016,Cho2019,Paik2017c,PVWadekar_2014,Nono-Tchiomo2019,YHe_2021,Lebens-Higgins2016a,RZhang_2021,TMurauskas_2022,JShiogai_2016,KEom_2021}. Unfortunately due in part to a lack of lattice-matched substrates,  La:BaSnO$_3$ films suffer from a high density of structural defects, stacking faults and point defects, which limit their electron mobility. Noteworthy structural defects in epitaxial La:BaSO$_3$  films are threading dislocations (TDs), the density of which are often in the order of  $1.5\times10^{11}\text{ cm}^{-2}$ and higher for these films~\cite{NonoTchiomo2022,Paik2017c,Prakash2017c,JShiogai_2016,Postiglione2021,JKang2022}. Such TDs are due to the large lattice mismatch between La:BaSO$_3$ films and 
commercially available substrates. As shown in Fig.~\ref{Fig_01}\textcolor{blue}{{(a)}}, the commercially available substrate with the closest lattice match is the scandate material PrScO$_3$, which presents compressive lattice of -2.3\%~\cite{TMGesing2009}. Other usual perovskite oxide substrates such as SrTiO$_3$, (LaAlO$_3$)$_{0.3}$(Sr$_2$AlTaO$_6$)$_{0.7}$ (LSAT) and LaAlO$_3$ have compressive lattice mismatches of -5.4\%, -6.4\%, and -8.6\%, respectively. Although the rock-salt substrates like MgO offer moderate tensile mismatch of +2.3\%, however they have a structure and symmetry mismatch to the La:BaSnO$_3$ perovskite structure. 

Furthermore, other factors such as complex point defects, Ba/Sn antisites and Ruddlesden–Popper shear faults that form during epitaxial growth are also known to act as extra electron traps or scattering sites, which limit electron mobility in films~\cite{Paik2017c,Nono-Tchiomo2019,ZWang_2019,WYWang2015,JShiogai_2016,Raghavan2016c}. Thus, as compared to bulk single crystals, the reported RT $\mu_e$ in epitaxial La:BaSnO$_3$ films have only reached a  maximum  value  of $183\text{ cm}^2\text{ V}^{-1}\text{s}^{-1}$ ($n\simeq 1.2 \times 10^{20} \text{ cm}^{-3} $) for films prepared by molecular beam epitaxy (MBE)~\cite{Paik2017c}. Other film deposition techniques achieved maximum RT $\mu_e$ of $140\text{ cm}^2\text{ V}^{-1}\text{s}^{-1}$ ($n\simeq 5.2 \times 10^{20} \text{ cm}^{-3} $) for pulsed laser deposition (PLD)~\cite{Nono-Tchiomo2019}, $121\text{ cm}^2\text{ V}^{-1}\text{s}^{-1}$ ($n \simeq 4.0\times 10^{20} \text{ cm}^{-3} $) for high-pressure magnetron sputtering~\cite{RZhang_2021}, $96\text{ cm}^2\text{ V}^{-1}\text{s}^{-1}$ ($n \simeq 2.6\times 10^{20} \text{ cm}^{-3} $) for vacuum-annealed films grown by high-pressure-oxygen sputter deposition~\cite{Postiglione2021}, and $53\text{ cm}^2\text{ V}^{-1}\text{s}^{-1}$ ($n \simeq 2.0\times 10^{20} \text{ cm}^{-3} $) for the chemical solution deposition technique~\cite{YHe_2021}.

To reduce the defect densities in epitaxial films, various approaches  have been explored. These include, for example, the use of insulating buffer layers (e.g.,  (Sr,Ba)SnO$_3$ and BaSnO$_3$)  inserted between the substrate and the active La:BaSnO$_3$ top layers to sever the large lattice mismatch~\cite{Paik2017c,Prakash2017c,JShiogai_2016,SJuyeon_2016, FKohei_2016}; a very high-temperature-grown insulating buffer layer to reduce the density of TDs~\cite{Nono-Tchiomo2019}; the use of flux-grown undoped BaSnO$_3$ (001) single crystals as substrates~\cite{LWoong-Jhae_2016}; as well as post-growth annealing processes~\cite{Cho2019,Yoon2018,LWoong-Jhae_2016} and  adsorption-controlled MBE growth for improved stoichiometry control~\cite{PJisung_2020,Raghavan2016c,Lebens-Higgins2016a,ZWang_2019,Paik2017c,APrakash_2017,KEom_2021}. These prior approaches suggest that there is still room for exploring other strategies to boost  $\mu_e$ in epitaxial La:BaSnO$_3$ films.

By the combination of PLD and MBE, we report an effective synthetic route that employs a high-temperature-grown buffer layer to boost RT $\mu_e$ in epitaxial La:BaSnO$_\text{3}$ films. The density of TDs is reduced considerably by first growing an insulating buffer layer of SrZrO$_\text{3}$ at a very high temperature using PLD, followed by the epitaxial growth of an overlaying La:BaSnO$_3$ active layer by MBE.  Besides the demonstration of the enhancement of electron mobility in epitaxial La:BaSnO$_\text{3}$/SrZrO$_3$  heterostructures prepared on TbScO$_3$~\cite{Nono-Tchiomo2019}, the current study reveals that the insertion of a very high-temperature-grown SrZrO$_3$ epitaxial layer between the film and the substrate is an effective synthetic route for minimizing the density of defects and boosting the transport properties of La:BaSnO$_3$ films prepared on most oxide substrates. We demonstrate that this synthesis approach is applicable to many oxide substrates that induce large compressive or tensile strains to the films, which is a significant contribution for addressing the challenge of lack of a commercially available lattice-matched substrates close to the BaSnO$_\text{3}$ cubic lattice parameter  (4.116\AA). The effectiveness of the synthesis approach has been explored by preparing La:BaSnO$_\text{3}$/SrZrO$_3$   heterostructures on scandate DyScO$_3$ and TbScO$_3$ substrates, and also on rock-salt MgO substrates. Surface and structural characterization demonstrates smooth surface morphologies and high crystalline quality of the films. Electronic transport measurements revealed RT $\mu_e$ as high as $157\text{ cm}^2\text{ V}^{-1}\text{s}^{-1}$ ($n \simeq 1.27\times 10^{20} \text{ cm}^{-3} $), $145\text{ cm}^2\text{ V}^{-1}\text{s}^{-1}$ ($n \simeq 1.13\times 10^{20} \text{ cm}^{-3} $), and $143\text{ cm}^2\text{ V}^{-1}\text{s}^{-1}$ ($n \simeq 1.57\times 10^{20} \text{ cm}^{-3} $) for heterostructures grown on DyScO$_3$, MgO, and TbScO$_3$ crystalline substrates, respectively. As compared to prior reports, these RT $\mu_e$ are  the second-highest mobilities achieved in epitaxial La:BaSnO$_3$ films; and so far, the highest RT $\mu_e$ obtained  for La:BaSnO$_3$ films  of  small  thickness  ($\leq 25$ nm) prepared using non-BaSnO$_3$ buffer layers.

\begin{figure*}[!t]
     \includegraphics[width=1\textwidth]{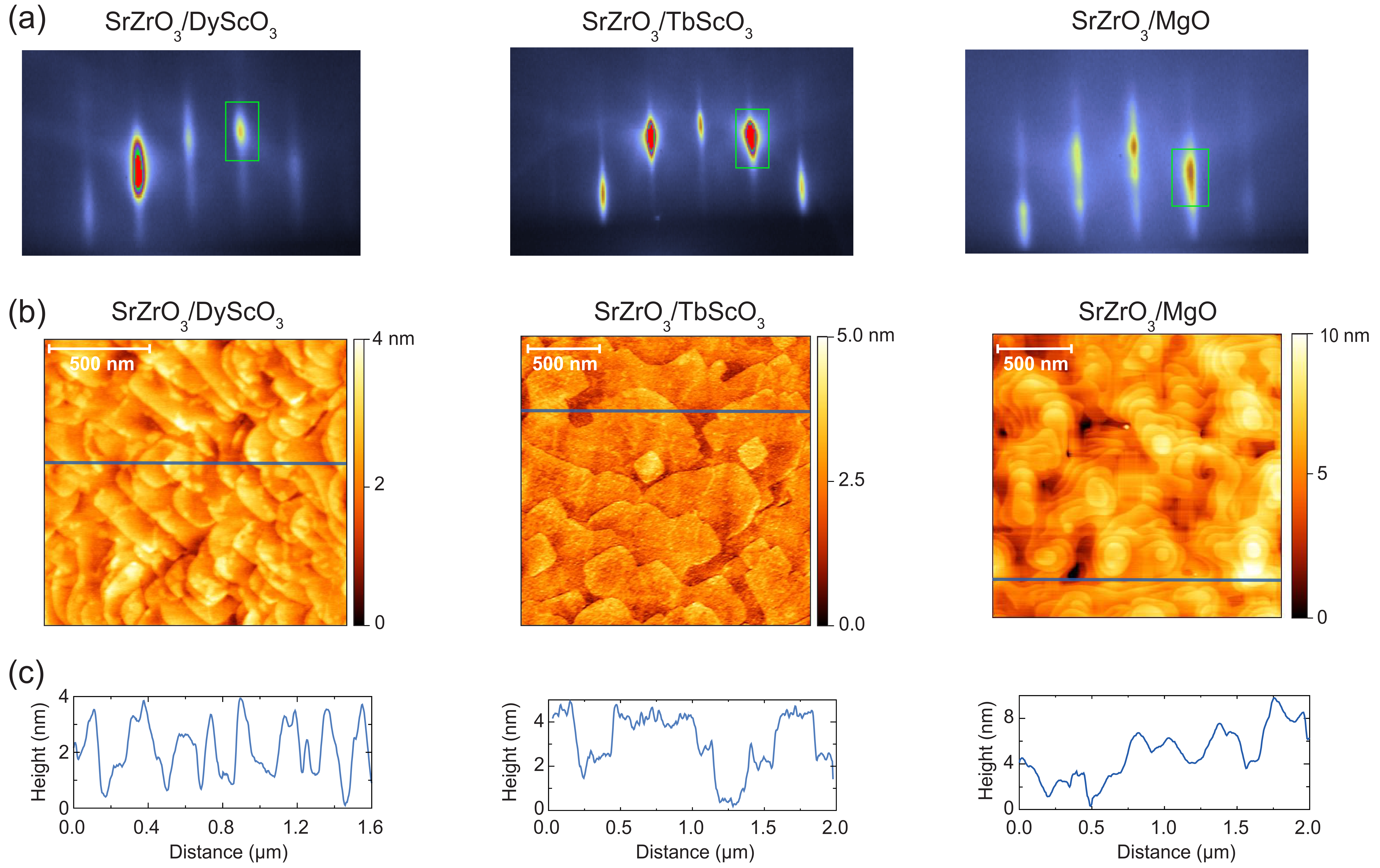}
    \caption{\small{(a) Reflection high-energy electron diffraction  patterns of 100-nm-thick SrZrO$_3$ buffer layers grown on $(110)$-oriented DyScO$_3$ and TbScO$_3$, and $(001)$-oriented MgO  crystalline substrates. The green rectangles mark the region from which the integrated intensity as a function of time during deposition were recorded [see Fig.~\textcolor{blue}{S1} of  the  supplementary  material]. (b) Atomic force microscopy images displaying the surface morphology, and (c) corresponding surface roughness profiles of the SrZrO$_3$ buffer layers on DyScO$_3$, TbScO$_3$, and MgO.}} 
\label{Fig_02}
\end{figure*}

Figure~\ref{Fig_01}\textcolor{blue}{(b)} depicts a schematic view of the sample types investigated. Our approach to minimizing dislocation density starts by growing at very high temperature (1300\degree C) an insulating SrZrO$_3$ buffer layer using PLD. The SrZrO$_3$ layers  were deposited on several (001)-oriented MgO,  (110)-oriented DyScO$_3$, and TbScO$_3$  crystalline  substrates  ($5\times 5 \times 1 \text{ mm}^3$). Prior to  deposition,  all the substrates were terminated \textit{in situ} at very high temperatures using  a  CO$_2$ laser  substrate  heating  system, as  described in Ref.~\cite{WBraun_2020}. To grow SrZrO$_3$ buffer layers by PLD ($\lambda=248$ nm), we used a laser  fluence  of $2 \text{ J cm}^{-2}$ at $1.4 \times 10^{-2} \text{ mbar}$ of  O$_2$. The buffer layers were deposited at 4 Hz to a thickness of 100 nm. SrZrO$_3$ is chosen because it has a low vapor pressure and can  therefore  be  grown  at  high  temperatures~\cite{Nono-Tchiomo2019}. Also, SrZrO$_3$ has a psuedocubic lattice parameter value (4.101 Å) that is very close to that of  La:BaSnO$_3$ [Fig.~\ref{Fig_01}\textcolor{blue}{(a)}]. Ideally, an undoped BaSnO$_\text{3}$ buffer layer grown at higher substrate temperatures could also be used to lower dislocation densities and improve $\mu_e$ further, but due to the significant volatility of tin oxide at substrate temperatures above $~ 850$\degree C, this is not a viable option. Details about the PLD growth of the SrZrO$_3$ buffer films are provided in Ref.~\cite{Nono-Tchiomo2019}.

Epitaxial La:BaSnO$_3$ (25 nm) films  were grown on top of the SrZrO$_3$ buffer layers using a Veeco GEN10 MBE system. Separate effusion cells containing lanthanum (99.996\% purity, Ames Lab), barium (99.99\% purity, Sigma-Aldrich), and SnO$_2$ (99.996\% purity, Alfa Aesar) were heated. The fluxes of the resulting molecular-beams emanating from the effusion cells were measured by a quartz crystal microbalance before growth. The La:BaSnO$_3$ films were grown in an adsorption-controlled regime by supplying an excess SnO$_{\text{x}}$-flux. The background pressure of the oxidant, $10\% \text{ O}_3+90\%\text{ O}_2$, was held at a constant ion gauge pressure of $1.33 \times 10^{-6} \text{ mbar}$. The substrate temperature was maintained between 830 and 850\degree C, as measured by an optical pyrometer. Details on the growth of  La:BaSnO$_3$ films by MBE are provided in Ref.~\cite{Paik2017c}.

For transport measurements, we used a Nanometrics Hall measurement system to characterize the resistivity, carrier concentration, $n$, and the electron mobility, $\mu_e$, of the La:BaSnO$_3$ films, using four spring-loaded tips (Au/Ir) arranged in a Van der Pauw geometry.
 
\begin{figure*}[!t]
     \includegraphics[width=1\textwidth]{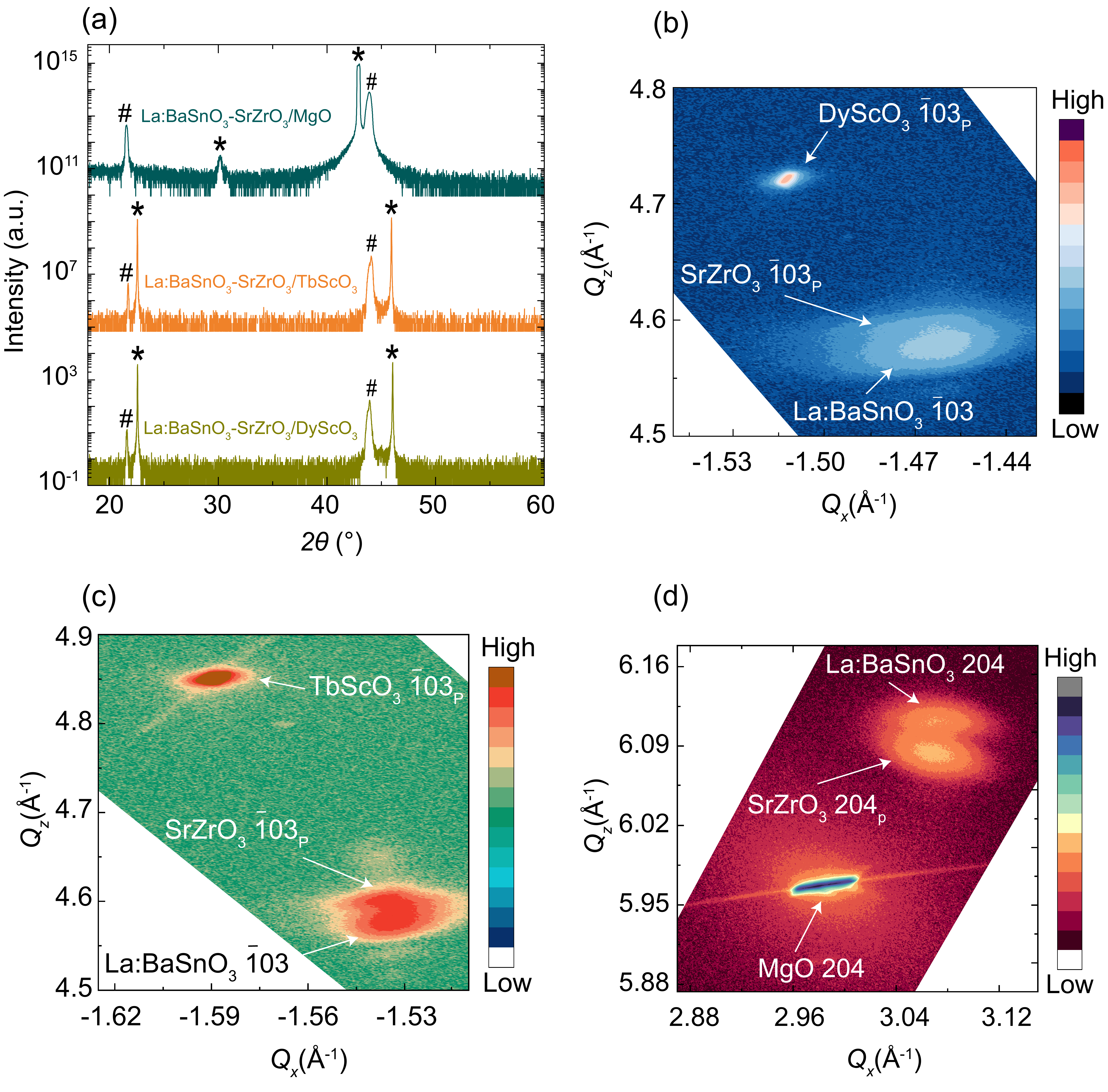}
    \caption{\small{Representative XRD scans of the La:BaSnO$_3$/SrZrO$_3$ heterostructures grown on different substrates. (a) $\theta-2\theta$ scans of the La:BaSnO$_3$/SrZrO$_3$ heterostructures grown on  DyScO$_3$, TbScO$_3$, and  MgO substrates. Only the substrate peaks ($\textbf{*}$) and the $0 0 l $ family of the films' diffraction peaks ($\textbf{\#}$) are resolved. Reciprocal space maps (RSM) of the films in (b) and (c) around the  $\bar{1}03_{\text{p}}$ reflection peaks of La:BaSnO$_3$/SrZrO$_3$, and around the $\bar{1}03_{\text{p}}$ reflection of DyScO$_3$ and TbScO$_3$ substrates, where $\text{p}$ refers to pseudocubic indices. (d) RSM around the $204_p$ reflection peak of the  La:BaSnO$_3$/SrZrO$_3$ film, and the $204$ reflection of the MgO substrate.}}
\label{Fig_03}
\end{figure*}
\begin{table*}[t]
	\caption{\label{Tab_01}Electronic transport characteristics (carrier density and mobility) of the samples discussed in this study.}
	\begin{ruledtabular}
		\begin{tabular}{cccc}
			Sample name & Sample layout & Carrier density & Carrier mobility \\
			& & ($\times \text{10}^{\text{20}}$~cm$^{-\text{3}}$) & (cm$^\text{2}$~V$^{-\text{1}}$~s$^{-\text{1}}$) \\
			\hline
			A & La:BaSnO$_\text{3}$ (25~nm)/SrZrO$_\text{3}$ (100~nm)/DyScO$_\text{3}$ & $\text{1.27}\pm\text{0.05}$ & $\text{157}\pm\text{2}$ \\
			B & La:BaSnO$_\text{3}$ (25~nm)/SrZrO$_\text{3}$ (100~nm)/TbScO$_\text{3}$ & $\text{1.57}\pm\text{0.05}$ & $\text{143}\pm\text{2}$ \\
			C & La:BaSnO$_\text{3}$ (25~nm)/SrZrO$_\text{3}$ (100~nm)/MgO & $\text{1.13}\pm\text{0.05}$ & $\text{145}\pm\text{2}$ \\
			D & La:BaSnO$_\text{3}$ (25~nm)/SrZrO$_\text{3}$ (100~nm)/DyScO$_\text{3}$ & $\text{1.47}\pm\text{0.05}$ & $\text{148}\pm\text{2}$ \\
			E & La:BaSnO$_\text{3}$ (25~nm)/SrZrO$_\text{3}$ (100~nm)/TbScO$_\text{3}$ & $\text{1.56}\pm\text{0.05}$ & $\text{143}\pm\text{2}$ \\
			F & La:BaSnO$_\text{3}$ (25~nm)/SrZrO$_\text{3}$ (100~nm)/MgO & $\text{1.06}\pm\text{0.05}$ & $\text{131}\pm\text{2}$ \\
			
		\end{tabular}
	\end{ruledtabular} 
\end{table*}
We  first  characterize the buffer layer. The high-temperature growth of the SrZrO$_3$ buffer  layers on different substrates was in situ monitored  by  reflection  high-energy electron diffraction (RHEED) [Fig.~\textcolor{blue}{S1(a)}-\textcolor{blue}{(c)}].
RHEED oscillations and sharp, diffracted and specular RHEED patterns  were observed throughout the deposition of the SrZrO$_3$ layers on DyScO$_3$, TbScO$_3$ and MgO substrates, indicating a smooth film surface [Fig.~\ref{Fig_02}\textcolor{blue}{(a)}]. The time-dependent RHEED intensity oscillations observed for SrZrO$_3$ buffer layers prepared on DyScO$_3$ and TbScO$_3$ substrates [Fig.~\textcolor{blue}{S1(a)}-\textcolor{blue}{(b)}] are suggestive of a layer-by-layer growth mode for these buffer layers prepared at very high-temperature. Based on these, we estimate the thickness of the SrZrO$_3$ intermediate layer to 100 nm, in consistency with the scanning transmission electron microscopy cross sections. The surface morphology of a SrZrO$_3$ layers was investigated using atomic force microscopy (AFM). Figure~\ref{Fig_02}\textcolor{blue}{(b)} depicts typical AFM images for representative 100 nm thick SrZrO$_3$ grown at 1300\degree C on DyScO$_3$, TbScO$_3$, and MgO substrates. From a surface morphology point of view on a small scale, all samples exhibit a relatively smooth surface. We observe slight variations of the surface morphology for samples grown on different substrates; in particular, the SrZrO$_3$ layers grown on MgO exhibit some island growth. SrZrO$_3$ films grown by PLD are known to exhibit a significant surface roughness~\cite{LSijun_2015}. The existence of the small islands in these SrZrO$_3$ buffer layers may be an indication of nucleation sites caused  by interfacial strain energy originating from the lattice mismatch between the film and substrates, which results in a slightly increased surface roughness. For a lateral scan size of $\sim 2\times 2\,\mu\text{m}^2$, the extracted surface roughness is around 4 nm for  100-nm-thick SrZrO$_3$ layers grown on scandate substrates (DyScO$_3$ and TbScO$_3$); and  it increases to $\sim 8$ nm for SrZrO$_3$ films grown on MgO. 
[Fig.~\ref{Fig_02}\textcolor{blue}{(c)}]. From
$\theta-2\theta$ x-ray diffraction (XRD) scans of the buffer layers, we extracted out-of-plane lattice parameters of \textit{c} $=4.11 \pm 0.02$ Å for the 100 nm-thick SrZrO$_3$ films grown on DyScO$_3$, TbScO$_3$, and MgO substrates. These values are within experimental error of the fully relaxed psuedocubic lattice constant of SrZrO$_3$, $4.101$ Å.  Fully relaxed films are expected given the 100 nm thickness of the SrZrO$_3$ buffer layer, the high growth temperature, and the significant (2.3\% to 4.1\%) lattice mismatch between SrZrO$_3$ and these substrates. 
\begin{figure}[!t]
     \includegraphics[width=0.49\textwidth]{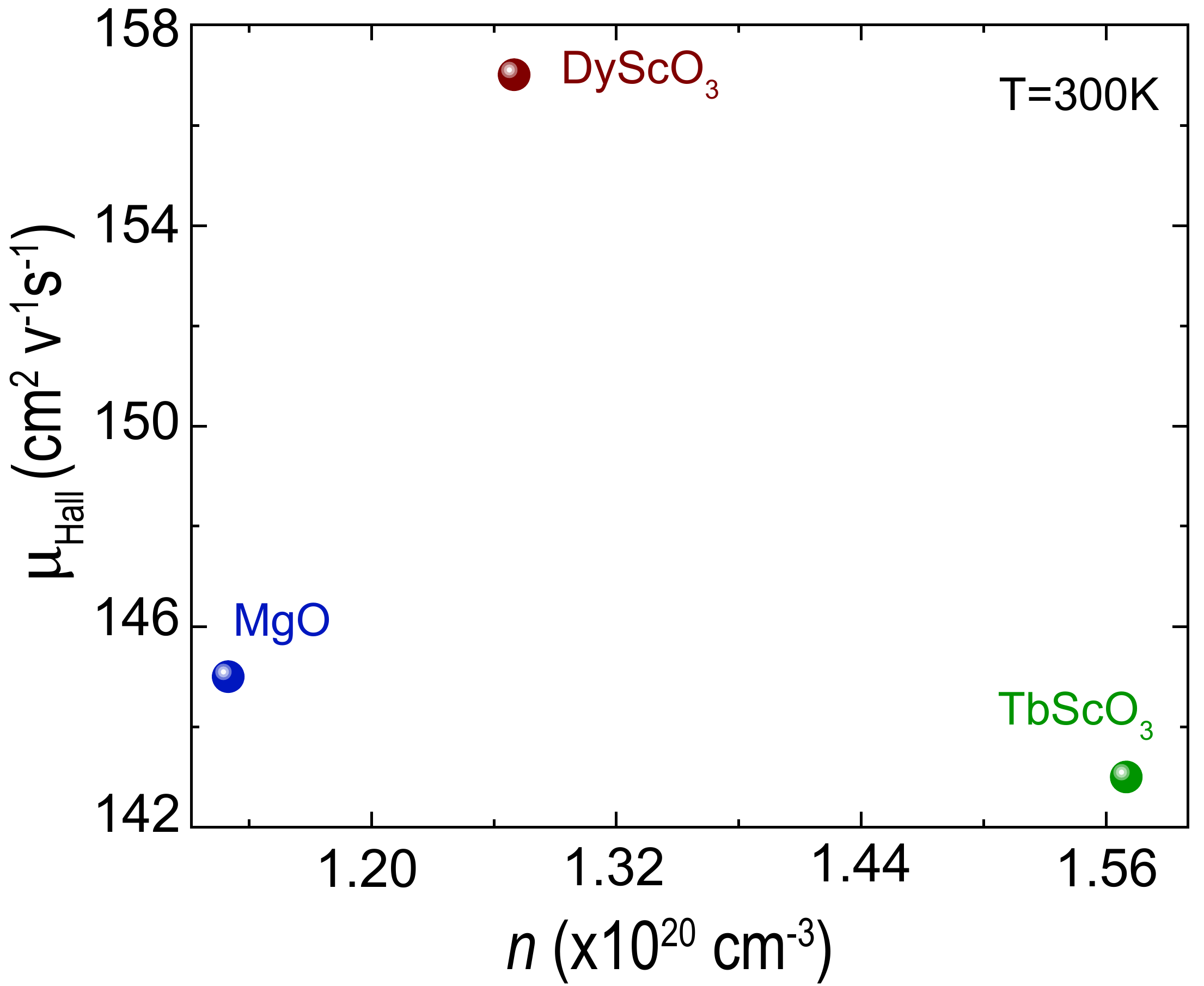}
    \caption{\small{Room-temperature electronic transport  characteristics, carrier density ($n$) and electron mobility ($\mu_e$), of representative La:BaSnO$_3$ (25 nm)/SrZrO$_3$ (100 nm) heterostructures grown on $(110)$-oriented DyScO$_3$, $(110)$-oriented TbScO$_3$, and $(001)$-oriented MgO  crystalline substrates.}} 
\label{Fig_04}
\end{figure}

Next we used adsorption-controlled MBE to deposit 25-nm-thick La:BaSnO$_3$ films on top of PLD-grown SrZrO$_3$ prepared on different oxide substrates. Results discussed below demonstrate that our approach to use very high-temperature-grown SrZrO$_3$ buffer layers to reduce the density of TDs is applicable not only for epitaxial heterostructures prepared in the same deposition chamber without exposing samples to ambient conditions as reported in Ref.~\cite{Nono-Tchiomo2019} but also for SrZrO$_3$ buffer layers exposed to air for days prior to the subsequent epitaxial growth of La:BaSnO$_3$ active layers~\cite{Note_01}.
\begin{figure*}[!t]
     \includegraphics[width=1\textwidth]{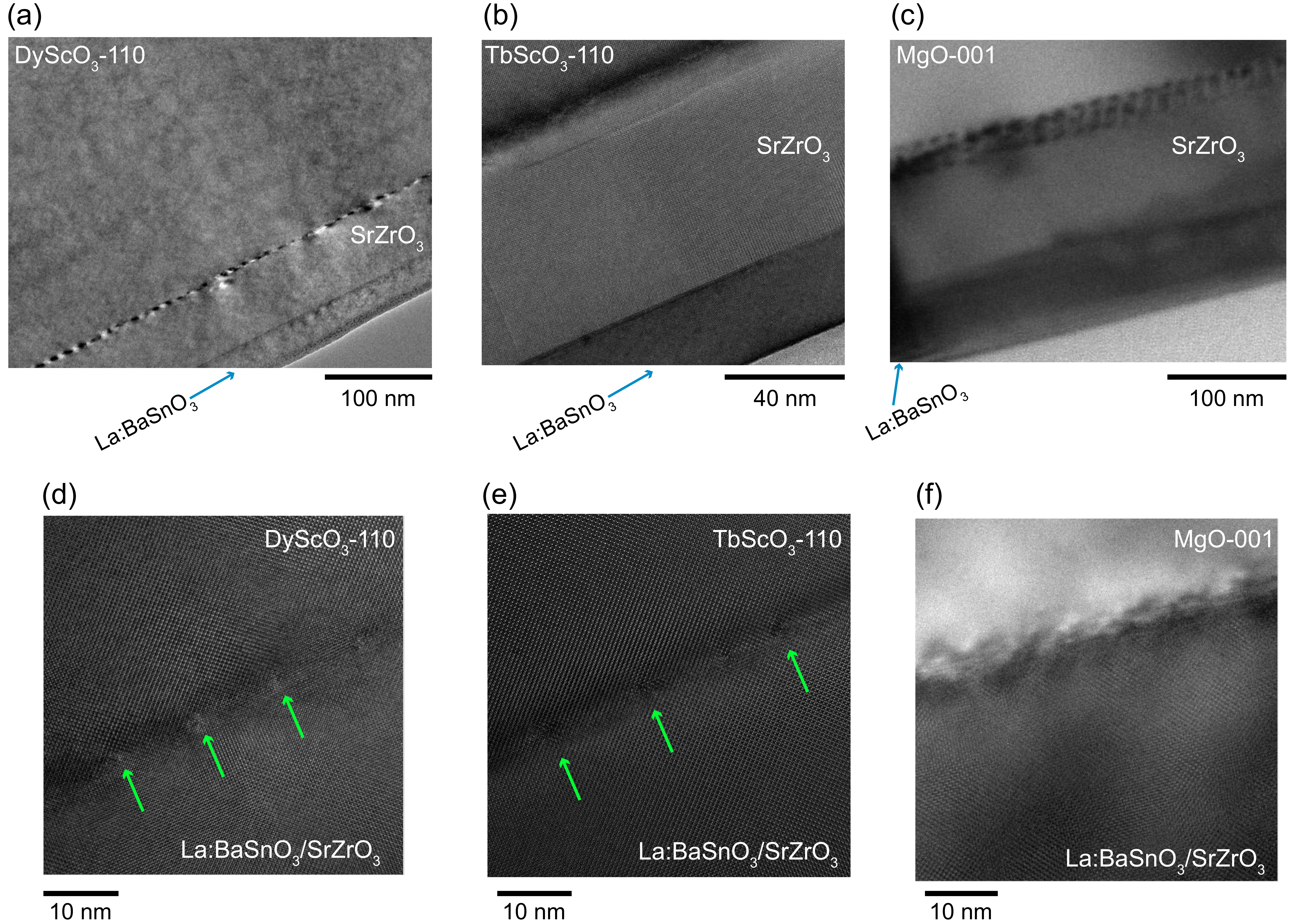}
    \caption{\small{Bright-field TEM images of representative  La:BaSnO$_3$/SrZrO$_3$ heterostructure grown on (a) $(110)$-oriented DyScO$_3$, (b)  $(110)$-oriented TbScO$_3$, and (c) $(001)$-oriented MgO  crystalline substrates. There are no discernible edge-type threading dislocations in these TEM images. Only misfit dislocations are visible along the interface between the films and substrates. High-resolution scanning transmission electron microscopy (HRSTEM) images for the same heterostructures grown on (d) DyScO$_3$, (e)  TbScO$_3$, and (f) MgO.  Few misfit dislocations are visible for the heterostructures grown on DyScO$_3$ and TbScO$_3$ substrates (indicated by green arrows). (f) For the film grown on MgO, a higher density of misfit dislocations at the interface (high strain) is observed.}} 
\label{Fig_05}
\end{figure*}

After MBE growth, the  crystalline  quality  and  phase  purity  of La:BaSnO$_3$/SrZrO$_3$ heterostructures  were characterized  by XRD. Figure~\ref{Fig_03}\textcolor{blue}{(a)} shows the representative $\theta-2\theta$ scans for the La:BaSnO$_3$/SrZrO$_3$/DyScO$_3$ (sample A), La:BaSnO$_3$/SrZrO$_3$/TbScO$_3$(sample B),  and  La:BaSnO$_3$/SrZrO$_3$/MgO (sample C) heterostructures. Only  the  substrate  peaks  and phase-pure 0  0  $l$  family  of  the film diffraction  peaks are resolved, indicating a high crystallinity; and also, verifying that the heterostructures were aligned along the c-axis. The extracted out-of-plane lattice parameter of all three films is \textit{c}$=4.13 \pm 0.01$ Å. This value is close to the bulk lattice constant of BaSnO$_3$, and consistent with the out-of-plane lattice constants reported previously in La:BaSnO$_3$ films~\cite{Nono-Tchiomo2019,HMun_2013,Paik2017c,LWoong-Jhae_2016}. Figure~\textcolor{blue}{S3} shows a closeup view of the $\theta-2\theta$ scan around the $002$ diffraction peak for the La:BaSnO$_3$/SrZrO$_3$ heterostructures grown on the three substrates. The asymmetry in the peaks highlight the presence of the La:BaSnO$_3$ and SrZrO$_3$ layers in these heterostructures. In particular, the 002 peaks exhibited by heterostructures grown on DyScO$_3$ and TbScO$_3$ substrates reveal noticeable thickness fringes. The observation of Laue thickness fringes and solely $00l$ peaks is an indication of phase purity and smooth growth.

Figures~\ref{Fig_03}\textcolor{blue}{(b)} and \ref{Fig_03}\textcolor{blue}{(c)} show  reciprocal  space  maps  (RSM)  around the asymmetric $\bar{1}03_p$ reflection peaks of the heterostructures (samples A and B) prepared on scandate substrates (DyScO$_3$ and TbScO$_3$). Figure~\ref{Fig_03}\textcolor{blue}{(d)} is for the $204_p$ reflection peak of the film (sample C) on MgO.  From all three RSM maps, it is evident that the La:BaSnO$_3$ layer is commensurately strained to the SrZrO$_3$ buffer layer, but in all cases that the commensurate La:BaSnO$_3$/SrZrO$_3$ bilayer is relaxed from the underyling substrate. This makes sense given the excellent lattice match and structural match between La:BaSnO$_3$ and the SrZrO$_3$ buffer layer. This result is also consistent with  the literature for epitaxial La:BaSnO$_3$/SrZrO$_3$ heterostructures~\cite{Nono-Tchiomo2019}.  

Now we turn to the electronic transport and microstructural data. Figure~\ref{Fig_04} presents the electron mobility at 300 K as a function of carrier density for samples A, B, and C. The highest RT $\mu_e$ of $157\text{ cm}^2\text{ V}^{-1}\text{s}^{-1}$ with a carrier concentration of $n \simeq 1.27\times 10^{20} \text{ cm}^{-3} $ is achieved for sample A. This  RT $\mu_e$ is $\sim 10\%$ higher than the previously reported mobility ($140\text{ cm}^2\text{ V}^{-1}\text{s}^{-1}$) achieved by inserting a high-temperature-grown SrZrO$_3$ buffer layer  between the La:BaSnO$_3$ film and the substrate~\cite{Nono-Tchiomo2019}. It is the second-highest reported RT $\mu_e$ achieved in epitaxial La:BaSnO$_3$ films and the highest attained for thin ($\leq 25$ nm) epitaxial La:BaSnO$_3$ films. For samples B and C, we achieve RT $\mu_e$ of $143\text{ cm}^2\text{ V}^{-1}\text{s}^{-1}$ ($n \simeq 1.57\times 10^{20} \text{ cm}^{-3} $) and $145\text{ cm}^2\text{ V}^{-1}\text{s}^{-1}$ ($n \simeq 1.13\times 10^{20} \text{ cm}^{-3} $), respectively. These RT $\mu_e$ are reproducible in different structurally comparable heterostructures prepared in similar conditions [Table~\ref{Tab_01}]. The observed slight carrier mobility difference in structurally comparable heterostructures (samples A and D, and sample C and F) is attributed to experimental fluctuations. Also, as the thickness of the active layer is thin, the interface defect density or reconstruction effects may vary in structurally comparable samples; thus,  causing the observed slight carrier mobility difference. The achieved improvements in RT $\mu_e$ of La:BaSnO$_3$ films on different oxide substrates are attributed to the use of the high-temperature-grown buffer layer, which is know to minimize the density of defects, and thus results in an increase in carrier mobility.

To investigate the defect density and provide complementary real-space structural characterization of these films, cross-sectional transmission  electron microscopy (TEM) imaging was performed.  Figure~\ref{Fig_05}\textcolor{blue}{(a)}, ~\ref{Fig_05}\textcolor{blue}{(b)}, and ~\ref{Fig_05}\textcolor{blue}{(c)} depict bright-field TEM images of the entire film thickness for representative (25 nm) La:BaSnO$_3$/(100 nm) SrZrO$_3$ heterostructures prepared on DyScO$_3$,   TbScO$_3$, and MgO substrates, respectively. We observe misfit dislocations along the interface between the films and substrates. As expected for high-temperature-grown SrZrO$_3$ buffer layers~\cite{Nono-Tchiomo2019}, TDs  were  barely observed in all the three representative samples [Fig.~\ref{Fig_05}\textcolor{blue}{(a)}-\textcolor{blue}{(c)}]. STEM investigations over wide areas showed hardly any TDs in La:BaSnO$_3$/SrZrO$_3$ heterostructures, and electron  energy-loss  spectroscopy (EELS) map analyses indicate expected elemental composition in the films [see Fig.~\textcolor{blue}{S2(a)}-\textcolor{blue}{(c)} of  the  supplementary  material]. For films prepared on DyScO$_3$,   TbScO$_3$, and MgO substrates, the extract upper limit of TD density is well below $1.0\times 10^{10} \text{ cm}^{-2}$, in agreement with previous report~\cite{Nono-Tchiomo2019}.

The  low  density  of  TDs  in  these samples  is  attributed  to  the very high temperature (1300 \degree C) used for the growth of the SrZrO$_3$ buffer layer. This approach helps to eliminate most of TDs that would act as scattering centers  and  trap  electrons. As SrZrO$_3$ has an excellent lattice match to La:BaSnO$_3$, inserting a high-temperature-grown SrZrO$_3$ layer between the La:BaSnO$_3$ film and substrate minimizes the TD density~\cite{Nono-Tchiomo2019}. At the high substrate temperature used for the growth of the SrZrO$_3$ buffer layer, which has a significant lattice mismatch to all of the underlying commercial substrates used, the TDs are able to more readily move and react with each other. The result is misfit dislocation segments that relieve the misfit strain and a lower TD density than would be the case had the SrZrO$_3$ buffer layer been grown at lower temperature where the TDs are less mobile. The lowered TD density of the SrZrO$_3$ directly benefits the subsequently grown La:BaSnO$_3$ layer as it is not only well lattice matched to the SrZrO$_3$ buffer layer, but inherits relatively few TDs from it. Notably for our growth approach, the subsequent growth of an overlying La:BaSnO$_3$ by adsorption-controlled MBE helps in achieving better stoichiometry control, thus allowing to enhance electron mobility in these films. Our results clearly demonstrate that the high-temperature-grown SrZrO$_3$ epilayer is a suitable  template for subsequent growth of high mobility La:BaSnO$_3$ films with fewer TDs not  only  on  TbScO$_3$~\cite{Nono-Tchiomo2019}, but  also  on  other  oxide substrates (DyScO$_3$ and  MgO). It is envisaged that this synthesis approach of high mobility La:BaSnO$_3$ films could also be extended to most, if not all, oxide substrates such as SrTiO$_3$, (LaAlO$_3$)$_{0.3}$(Sr$_2$AlTaO$_6$)$_{0.7}$ (LSAT),  LaAlO$_3$, MgAl$_2$O$_4$, and LaLuO$_3$ that present large compressive or tensile lattice mismatches to La:BaSnO$_3$.

Although our improved synthesis approach of epitaxial La:BaSnO$_3$/SrZrO$_3$ heterostructures reduces TDs and increases RT mobilities, we are not able to achieve reported mobilites as high as those reported in MBE grown thick (60 nm) La:BaSnO$_3$/ (330 nm) BaSnO$_3$ films~\cite{Paik2017c}. As the La:BaSnO$_3$ active  layer  is  thin  (25  nm) in our heterostructures,  it could be that not only TDs are trapping electrons, but also effects such as surface scattering or interface traps are lowering the density of mobile carriers. These contributions are expected to be  less pronounced in thick La:BaSnO$_3$/BaSnO$_3$ heterostructures as the buffer layer and the active layer consist of almost the same materials.

In summary, we have explored an approach to enhance room-temperature electron mobility in La:BaSnO$_3$/SrZrO$_3$ heterostructures. For MBE-grown La:BaSnO$_3$ films  prepared  on  PLD-grown SrZrO$_3$ buffer  layers that were grown at 1300\degree C,   we achieve RT mobilities  of  157, 145, and 143  cm$^2$V$^{-1}$s$^{-1}$ for films prepared on DyScO$_3$, MgO and TbScO$_3$ substrates, respectively.  The density of TDs are very low in these films with an upper limit well below $ 1.0\times 10^{10} \text{ cm}^{-2}$ for all films prepared on these oxide substrates, thus verifying the efficacy of our synthesis approach. Our work provides an effective approach for the growth of high mobility La:BaSnO$_3$ epitaxial films on most, if not all,  oxide substrates that present large compressive or tensile lattice mismatches to La:BaSnO$_3$, which is an essential step in tackling the challenges caused by the lack of commercially available substrates with lattice parameters matching the BaSnO$_3$  unit cell. Also, we note that achieving high RT $\mu_e$ at low thickness and relatively low carrier concentrations in these heterostructures provides an opportunity to fabricate La:BaSnO$_3$-based FETs on various oxide substrates in which channels may be fully depleted. Based on these results, future directions are expected to focus on exploring the physics of La:BaSnO$_3$-based devices for their potential practical applications in oxide electronics. 
\\ \\
\indent See   supplementary   material   for additional surface and microstructural characterizations (RHEED, TEM and XRD) of the La:BaSnO$_3$/SrZrO$_3$ heterostructures.
\\ \\
P. Ngabonziza acknowledges startup funding from the  College of Science and the Department of Physics \& Astronomy at Louisiana State University. \newline \indent W. Sigle and P. van Aken  acknowledge  funding  from  the  European  Union’s  Horizon2020 research and innovation program under Grant Agreement No.823717-ESTEEM3. \newline \indent D. G. Schlom and J. Park acknowledge support by the Air Force Office of Scientific Research under Award No. FA9550-16-1-0192 and  gratefully acknowledge Professors Grace Xing and Debdeep Jena for use of their Nanometrics Hall measurement system. 
\\ 
\indent The authors declare no conflict of interest.
 \\ 
\indent The  data  that  support  the  findings  of  this  study  are  available from the corresponding authors upon reasonable request.
\bibliography{references-2023}

\begin{thebibliography}{41}%
\makeatletter
\providecommand \@ifxundefined [1]{%
 \@ifx{#1\undefined}
}%
\providecommand \@ifnum [1]{%
 \ifnum #1\expandafter \@firstoftwo
 \else \expandafter \@secondoftwo
 \fi
}%
\providecommand \@ifx [1]{%
 \ifx #1\expandafter \@firstoftwo
 \else \expandafter \@secondoftwo
 \fi
}%
\providecommand \natexlab [1]{#1}%
\providecommand \enquote  [1]{``#1''}%
\providecommand \bibnamefont  [1]{#1}%
\providecommand \bibfnamefont [1]{#1}%
\providecommand \citenamefont [1]{#1}%
\providecommand \href@noop [0]{\@secondoftwo}%
\providecommand \href [0]{\begingroup \@sanitize@url \@href}%
\providecommand \@href[1]{\@@startlink{#1}\@@href}%
\providecommand \@@href[1]{\endgroup#1\@@endlink}%
\providecommand \@sanitize@url [0]{\catcode `\\12\catcode `\$12\catcode
  `\&12\catcode `\#12\catcode `\^12\catcode `\_12\catcode `\%12\relax}%
\providecommand \@@startlink[1]{}%
\providecommand \@@endlink[0]{}%
\providecommand \url  [0]{\begingroup\@sanitize@url \@url }%
\providecommand \@url [1]{\endgroup\@href {#1}{\urlprefix }}%
\providecommand \urlprefix  [0]{URL }%
\providecommand \Eprint [0]{\href }%
\providecommand \doibase [0]{https://doi.org/}%
\providecommand \selectlanguage [0]{\@gobble}%
\providecommand \bibinfo  [0]{\@secondoftwo}%
\providecommand \bibfield  [0]{\@secondoftwo}%
\providecommand \translation [1]{[#1]}%
\providecommand \BibitemOpen [0]{}%
\providecommand \bibitemStop [0]{}%
\providecommand \bibitemNoStop [0]{.\EOS\space}%
\providecommand \EOS [0]{\spacefactor3000\relax}%
\providecommand \BibitemShut  [1]{\csname bibitem#1\endcsname}%
\let\auto@bib@innerbib\@empty
\bibitem [{\citenamefont {Kim}\ \emph {et~al.}(2012{\natexlab{a}})\citenamefont
  {Kim}, \citenamefont {Kim}, \citenamefont {Kim}, \citenamefont {Kim},
  \citenamefont {Mun}, \citenamefont {Jeon}, \citenamefont {Hong},
  \citenamefont {Lee}, \citenamefont {Ju}, \citenamefont {Kim},\ and\
  \citenamefont {Char}}]{Kim2012b}%
  \BibitemOpen
  \bibfield  {author} {\bibinfo {author} {\bibfnamefont {H.~J.}\ \bibnamefont
  {Kim}}, \bibinfo {author} {\bibfnamefont {U.}~\bibnamefont {Kim}}, \bibinfo
  {author} {\bibfnamefont {H.~M.}\ \bibnamefont {Kim}}, \bibinfo {author}
  {\bibfnamefont {T.~H.}\ \bibnamefont {Kim}}, \bibinfo {author} {\bibfnamefont
  {H.~S.}\ \bibnamefont {Mun}}, \bibinfo {author} {\bibfnamefont {B.-G.}\
  \bibnamefont {Jeon}}, \bibinfo {author} {\bibfnamefont {K.~T.}\ \bibnamefont
  {Hong}}, \bibinfo {author} {\bibfnamefont {W.-J.}\ \bibnamefont {Lee}},
  \bibinfo {author} {\bibfnamefont {C.}~\bibnamefont {Ju}}, \bibinfo {author}
  {\bibfnamefont {K.~H.}\ \bibnamefont {Kim}},\ and\ \bibinfo {author}
  {\bibfnamefont {K.}~\bibnamefont {Char}},\ }\href
  {https://doi.org/10.1143/apex.5.061102} {\bibfield  {journal} {\bibinfo
  {journal} {Appl. Phys. Express}\ }\textbf {\bibinfo {volume} {5}},\ \bibinfo
  {pages} {061102} (\bibinfo {year} {2012}{\natexlab{a}})}\BibitemShut
  {NoStop}%
\bibitem [{\citenamefont {Kim}\ \emph {et~al.}(2012{\natexlab{b}})\citenamefont
  {Kim}, \citenamefont {Kim}, \citenamefont {Kim}, \citenamefont {Kim},
  \citenamefont {Kim}, \citenamefont {Jeon}, \citenamefont {Lee}, \citenamefont
  {Mun}, \citenamefont {Hong}, \citenamefont {Yu}, \citenamefont {Char},\ and\
  \citenamefont {Kim}}]{Kim2012a}%
  \BibitemOpen
  \bibfield  {author} {\bibinfo {author} {\bibfnamefont {H.~J.}\ \bibnamefont
  {Kim}}, \bibinfo {author} {\bibfnamefont {U.}~\bibnamefont {Kim}}, \bibinfo
  {author} {\bibfnamefont {T.~H.}\ \bibnamefont {Kim}}, \bibinfo {author}
  {\bibfnamefont {J.}~\bibnamefont {Kim}}, \bibinfo {author} {\bibfnamefont
  {H.~M.}\ \bibnamefont {Kim}}, \bibinfo {author} {\bibfnamefont {B.-G.}\
  \bibnamefont {Jeon}}, \bibinfo {author} {\bibfnamefont {W.-J.}\ \bibnamefont
  {Lee}}, \bibinfo {author} {\bibfnamefont {H.~S.}\ \bibnamefont {Mun}},
  \bibinfo {author} {\bibfnamefont {K.~T.}\ \bibnamefont {Hong}}, \bibinfo
  {author} {\bibfnamefont {J.}~\bibnamefont {Yu}}, \bibinfo {author}
  {\bibfnamefont {K.}~\bibnamefont {Char}},\ and\ \bibinfo {author}
  {\bibfnamefont {K.~H.}\ \bibnamefont {Kim}},\ }\href
  {https://doi.org/10.1103/PhysRevB.86.165205} {\bibfield  {journal} {\bibinfo
  {journal} {Phys. Rev. B}\ }\textbf {\bibinfo {volume} {86}},\ \bibinfo
  {pages} {165205} (\bibinfo {year} {2012}{\natexlab{b}})}\BibitemShut
  {NoStop}%
\bibitem [{\citenamefont {Fujiwara}\ \emph {et~al.}(2017)\citenamefont
  {Fujiwara}, \citenamefont {Nishihara}, \citenamefont {Shiogai},\ and\
  \citenamefont {Tsukazaki}}]{KFujiwara_2017}%
  \BibitemOpen
  \bibfield  {author} {\bibinfo {author} {\bibfnamefont {K.}~\bibnamefont
  {Fujiwara}}, \bibinfo {author} {\bibfnamefont {K.}~\bibnamefont {Nishihara}},
  \bibinfo {author} {\bibfnamefont {J.}~\bibnamefont {Shiogai}},\ and\ \bibinfo
  {author} {\bibfnamefont {A.}~\bibnamefont {Tsukazaki}},\ }\href
  {https://doi.org/10.1063/1.4983611} {\bibfield  {journal} {\bibinfo
  {journal} {Appl. Phys. Lett.}\ }\textbf {\bibinfo {volume} {110}},\ \bibinfo
  {pages} {203503} (\bibinfo {year} {2017})}\BibitemShut {NoStop}%
\bibitem [{\citenamefont {Lee}\ \emph {et~al.}(2017)\citenamefont {Lee},
  \citenamefont {Kim}, \citenamefont {Kang}, \citenamefont {Jang},
  \citenamefont {Kim}, \citenamefont {Lee},\ and\ \citenamefont
  {Kim}}]{Lee2017}%
  \BibitemOpen
  \bibfield  {author} {\bibinfo {author} {\bibfnamefont {W.-J.}\ \bibnamefont
  {Lee}}, \bibinfo {author} {\bibfnamefont {H.~J.}\ \bibnamefont {Kim}},
  \bibinfo {author} {\bibfnamefont {J.}~\bibnamefont {Kang}}, \bibinfo {author}
  {\bibfnamefont {D.~H.}\ \bibnamefont {Jang}}, \bibinfo {author}
  {\bibfnamefont {T.~H.}\ \bibnamefont {Kim}}, \bibinfo {author} {\bibfnamefont
  {J.~H.}\ \bibnamefont {Lee}},\ and\ \bibinfo {author} {\bibfnamefont {K.~H.}\
  \bibnamefont {Kim}},\ }\href
  {https://doi.org/10.1146/annurev-matsci-070616-124109} {\bibfield  {journal}
  {\bibinfo  {journal} {Annu. Rev. Mater. Res.}\ }\textbf {\bibinfo {volume}
  {47}},\ \bibinfo {pages} {391} (\bibinfo {year} {2017})},\ \Eprint
  {https://arxiv.org/abs/and references therein} {and references therein}
  \BibitemShut {NoStop}%
\bibitem [{\citenamefont {Krishnaswamy}\ \emph {et~al.}(2016)\citenamefont
  {Krishnaswamy}, \citenamefont {Bjaalie}, \citenamefont {Himmetoglu},
  \citenamefont {Janotti}, \citenamefont {Gordon},\ and\ \citenamefont {Van~de
  Walle}}]{Krishnaswamy2016}%
  \BibitemOpen
  \bibfield  {author} {\bibinfo {author} {\bibfnamefont {K.}~\bibnamefont
  {Krishnaswamy}}, \bibinfo {author} {\bibfnamefont {L.}~\bibnamefont
  {Bjaalie}}, \bibinfo {author} {\bibfnamefont {B.}~\bibnamefont {Himmetoglu}},
  \bibinfo {author} {\bibfnamefont {A.}~\bibnamefont {Janotti}}, \bibinfo
  {author} {\bibfnamefont {L.}~\bibnamefont {Gordon}},\ and\ \bibinfo {author}
  {\bibfnamefont {C.~G.}\ \bibnamefont {Van~de Walle}},\ }\href
  {https://doi.org/10.1063/1.4942366} {\bibfield  {journal} {\bibinfo
  {journal} {Appl. Phys. Lett.}\ }\textbf {\bibinfo {volume} {108}},\ \bibinfo
  {pages} {083501} (\bibinfo {year} {2016})}\BibitemShut {NoStop}%
\bibitem [{\citenamefont {Naamneh}\ \emph {et~al.}(2022)\citenamefont
  {Naamneh}, \citenamefont {Guedes}, \citenamefont {Prakash}, \citenamefont
  {Cardoso}, \citenamefont {Shi}, \citenamefont {Plumb}, \citenamefont {Brito},
  \citenamefont {Jalan},\ and\ \citenamefont {Radović}}]{Naamneh2022}%
  \BibitemOpen
  \bibfield  {author} {\bibinfo {author} {\bibfnamefont {M.}~\bibnamefont
  {Naamneh}}, \bibinfo {author} {\bibfnamefont {E.~B.}\ \bibnamefont {Guedes}},
  \bibinfo {author} {\bibfnamefont {A.}~\bibnamefont {Prakash}}, \bibinfo
  {author} {\bibfnamefont {H.~M.}\ \bibnamefont {Cardoso}}, \bibinfo {author}
  {\bibfnamefont {M.}~\bibnamefont {Shi}}, \bibinfo {author} {\bibfnamefont
  {N.~C.}\ \bibnamefont {Plumb}}, \bibinfo {author} {\bibfnamefont {W.~H.}\
  \bibnamefont {Brito}}, \bibinfo {author} {\bibfnamefont {B.}~\bibnamefont
  {Jalan}},\ and\ \bibinfo {author} {\bibfnamefont {M.}~\bibnamefont
  {Radović}},\ }\href {https://doi.org/10.1038/s42005-022-01091-y} {\bibfield
  {journal} {\bibinfo  {journal} {Commun. Phys.}\ }\textbf {\bibinfo {volume}
  {5}},\ \bibinfo {pages} {317} (\bibinfo {year} {2022})}\BibitemShut {NoStop}%
\bibitem [{\citenamefont {Kim}\ \emph {et~al.}(2015)\citenamefont {Kim},
  \citenamefont {Park}, \citenamefont {Ha}, \citenamefont {Kim}, \citenamefont
  {Kim}, \citenamefont {Ju}, \citenamefont {Park}, \citenamefont {Yu},
  \citenamefont {Kim},\ and\ \citenamefont {Char}}]{SKim_2015}%
  \BibitemOpen
  \bibfield  {author} {\bibinfo {author} {\bibfnamefont {U.}~\bibnamefont
  {Kim}}, \bibinfo {author} {\bibfnamefont {C.}~\bibnamefont {Park}}, \bibinfo
  {author} {\bibfnamefont {T.}~\bibnamefont {Ha}}, \bibinfo {author}
  {\bibfnamefont {Y.~M.}\ \bibnamefont {Kim}}, \bibinfo {author} {\bibfnamefont
  {N.}~\bibnamefont {Kim}}, \bibinfo {author} {\bibfnamefont {C.}~\bibnamefont
  {Ju}}, \bibinfo {author} {\bibfnamefont {J.}~\bibnamefont {Park}}, \bibinfo
  {author} {\bibfnamefont {J.}~\bibnamefont {Yu}}, \bibinfo {author}
  {\bibfnamefont {J.~H.}\ \bibnamefont {Kim}},\ and\ \bibinfo {author}
  {\bibfnamefont {K.}~\bibnamefont {Char}},\ }\href
  {https://doi.org/10.1063/1.4913587} {\bibfield  {journal} {\bibinfo
  {journal} {APL Mater.}\ }\textbf {\bibinfo {volume} {3}},\ \bibinfo {pages}
  {036101} (\bibinfo {year} {2015})}\BibitemShut {NoStop}%
\bibitem [{\citenamefont {Yue}\ \emph {et~al.}(2018)\citenamefont {Yue},
  \citenamefont {Prakash}, \citenamefont {Robbins}, \citenamefont {Koester},\
  and\ \citenamefont {Jalan}}]{JYue_2018}%
  \BibitemOpen
  \bibfield  {author} {\bibinfo {author} {\bibfnamefont {J.}~\bibnamefont
  {Yue}}, \bibinfo {author} {\bibfnamefont {A.}~\bibnamefont {Prakash}},
  \bibinfo {author} {\bibfnamefont {M.~C.}\ \bibnamefont {Robbins}}, \bibinfo
  {author} {\bibfnamefont {S.~J.}\ \bibnamefont {Koester}},\ and\ \bibinfo
  {author} {\bibfnamefont {B.}~\bibnamefont {Jalan}},\ }\href
  {https://doi.org/10.1021/acsami.8b05229} {\bibfield  {journal} {\bibinfo
  {journal} {ACS Appl. Mater. Interfaces}\ }\textbf {\bibinfo {volume} {10}},\
  \bibinfo {pages} {21061} (\bibinfo {year} {2018})}\BibitemShut {NoStop}%
\bibitem [{\citenamefont {Cheng}\ \emph {et~al.}(2021)\citenamefont {Cheng},
  \citenamefont {Yang}, \citenamefont {Combs}, \citenamefont {Wu},
  \citenamefont {Kim}, \citenamefont {Chandrasekar}, \citenamefont {Wang},
  \citenamefont {Rajan}, \citenamefont {Stemmer},\ and\ \citenamefont
  {Lu}}]{JCheng_2021}%
  \BibitemOpen
  \bibfield  {author} {\bibinfo {author} {\bibfnamefont {J.}~\bibnamefont
  {Cheng}}, \bibinfo {author} {\bibfnamefont {H.}~\bibnamefont {Yang}},
  \bibinfo {author} {\bibfnamefont {N.~G.}\ \bibnamefont {Combs}}, \bibinfo
  {author} {\bibfnamefont {W.}~\bibnamefont {Wu}}, \bibinfo {author}
  {\bibfnamefont {H.}~\bibnamefont {Kim}}, \bibinfo {author} {\bibfnamefont
  {H.}~\bibnamefont {Chandrasekar}}, \bibinfo {author} {\bibfnamefont
  {C.}~\bibnamefont {Wang}}, \bibinfo {author} {\bibfnamefont {S.}~\bibnamefont
  {Rajan}}, \bibinfo {author} {\bibfnamefont {S.}~\bibnamefont {Stemmer}},\
  and\ \bibinfo {author} {\bibfnamefont {W.}~\bibnamefont {Lu}},\ }\href
  {https://doi.org/10.1063/5.0022550} {\bibfield  {journal} {\bibinfo
  {journal} {Appl. Phys. Lett.}\ }\textbf {\bibinfo {volume} {118}},\ \bibinfo
  {pages} {042105} (\bibinfo {year} {2021})}\BibitemShut {NoStop}%
\bibitem [{\citenamefont {Bucur}\ \emph {et~al.}(2012)\citenamefont {Bucur},
  \citenamefont {Bucur}, \citenamefont {Novaconi},\ and\ \citenamefont
  {Nicoara}}]{RABucur_2012}%
  \BibitemOpen
  \bibfield  {author} {\bibinfo {author} {\bibfnamefont {R.~A.}\ \bibnamefont
  {Bucur}}, \bibinfo {author} {\bibfnamefont {A.~I.}\ \bibnamefont {Bucur}},
  \bibinfo {author} {\bibfnamefont {S.}~\bibnamefont {Novaconi}},\ and\
  \bibinfo {author} {\bibfnamefont {I.}~\bibnamefont {Nicoara}},\ }\href
  {https://doi.org/https://doi.org/10.1016/j.jallcom.2012.07.072} {\bibfield
  {journal} {\bibinfo  {journal} {Journal of Alloys and Compounds}\ }\textbf
  {\bibinfo {volume} {542}},\ \bibinfo {pages} {142} (\bibinfo {year}
  {2012})}\BibitemShut {NoStop}%
\bibitem [{\citenamefont {Park}\ \emph {et~al.}(2020)\citenamefont {Park},
  \citenamefont {Paik}, \citenamefont {Nomoto}, \citenamefont {Lee},
  \citenamefont {Park}, \citenamefont {Grisafe}, \citenamefont {Wang},
  \citenamefont {Salahuddin}, \citenamefont {Datta}, \citenamefont {Kim},
  \citenamefont {Jena}, \citenamefont {Xing},\ and\ \citenamefont
  {Schlom}}]{PJisung_2020}%
  \BibitemOpen
  \bibfield  {author} {\bibinfo {author} {\bibfnamefont {J.}~\bibnamefont
  {Park}}, \bibinfo {author} {\bibfnamefont {H.}~\bibnamefont {Paik}}, \bibinfo
  {author} {\bibfnamefont {K.}~\bibnamefont {Nomoto}}, \bibinfo {author}
  {\bibfnamefont {K.}~\bibnamefont {Lee}}, \bibinfo {author} {\bibfnamefont
  {B.-E.}\ \bibnamefont {Park}}, \bibinfo {author} {\bibfnamefont
  {B.}~\bibnamefont {Grisafe}}, \bibinfo {author} {\bibfnamefont {L.-C.}\
  \bibnamefont {Wang}}, \bibinfo {author} {\bibfnamefont {S.}~\bibnamefont
  {Salahuddin}}, \bibinfo {author} {\bibfnamefont {S.}~\bibnamefont {Datta}},
  \bibinfo {author} {\bibfnamefont {Y.}~\bibnamefont {Kim}}, \bibinfo {author}
  {\bibfnamefont {D.}~\bibnamefont {Jena}}, \bibinfo {author} {\bibfnamefont
  {H.~G.}\ \bibnamefont {Xing}},\ and\ \bibinfo {author} {\bibfnamefont
  {D.~G.}\ \bibnamefont {Schlom}},\ }\href {https://doi.org/10.1063/1.5133745}
  {\bibfield  {journal} {\bibinfo  {journal} {APL Mater.}\ }\textbf {\bibinfo
  {volume} {8}},\ \bibinfo {pages} {011110} (\bibinfo {year}
  {2020})}\BibitemShut {NoStop}%
\bibitem [{\citenamefont {Nono~Tchiomo}\ \emph {et~al.}(2022)\citenamefont
  {Nono~Tchiomo}, \citenamefont {Carleschi}, \citenamefont {Prinsloo},
  \citenamefont {Sigle}, \citenamefont {van Aken}, \citenamefont {Mannhart},
  \citenamefont {Ngabonziza},\ and\ \citenamefont {Doyle}}]{NonoTchiomo2022}%
  \BibitemOpen
  \bibfield  {author} {\bibinfo {author} {\bibfnamefont {A.~P.}\ \bibnamefont
  {Nono~Tchiomo}}, \bibinfo {author} {\bibfnamefont {E.}~\bibnamefont
  {Carleschi}}, \bibinfo {author} {\bibfnamefont {A.~R.~E.}\ \bibnamefont
  {Prinsloo}}, \bibinfo {author} {\bibfnamefont {W.}~\bibnamefont {Sigle}},
  \bibinfo {author} {\bibfnamefont {P.~A.}\ \bibnamefont {van Aken}}, \bibinfo
  {author} {\bibfnamefont {J.}~\bibnamefont {Mannhart}}, \bibinfo {author}
  {\bibfnamefont {P.}~\bibnamefont {Ngabonziza}},\ and\ \bibinfo {author}
  {\bibfnamefont {B.~P.}\ \bibnamefont {Doyle}},\ }\href
  {https://doi.org/10.1063/5.0105116} {\bibfield  {journal} {\bibinfo
  {journal} {AIP Advances}\ }\textbf {\bibinfo {volume} {12}},\ \bibinfo
  {pages} {105019} (\bibinfo {year} {2022})}\BibitemShut {NoStop}%
\bibitem [{\citenamefont {Wang}\ \emph {et~al.}(2019)\citenamefont {Wang},
  \citenamefont {Paik}, \citenamefont {Chen}, \citenamefont {Muller},\ and\
  \citenamefont {Schlom}}]{ZWang_2019}%
  \BibitemOpen
  \bibfield  {author} {\bibinfo {author} {\bibfnamefont {Z.}~\bibnamefont
  {Wang}}, \bibinfo {author} {\bibfnamefont {H.}~\bibnamefont {Paik}}, \bibinfo
  {author} {\bibfnamefont {Z.}~\bibnamefont {Chen}}, \bibinfo {author}
  {\bibfnamefont {D.~A.}\ \bibnamefont {Muller}},\ and\ \bibinfo {author}
  {\bibfnamefont {D.~G.}\ \bibnamefont {Schlom}},\ }\href
  {https://doi.org/10.1063/1.5054810} {\bibfield  {journal} {\bibinfo
  {journal} {APL Mater.}\ }\textbf {\bibinfo {volume} {7}},\ \bibinfo {pages}
  {022520} (\bibinfo {year} {2019})}\BibitemShut {NoStop}%
\bibitem [{\citenamefont {Cho}\ \emph {et~al.}(2019{\natexlab{a}})\citenamefont
  {Cho}, \citenamefont {Feng}, \citenamefont {Onozato}, \citenamefont {Wei},
  \citenamefont {Sanchela}, \citenamefont {Ikuhara},\ and\ \citenamefont
  {Ohta}}]{HCho_2019}%
  \BibitemOpen
  \bibfield  {author} {\bibinfo {author} {\bibfnamefont {H.~J.}\ \bibnamefont
  {Cho}}, \bibinfo {author} {\bibfnamefont {B.}~\bibnamefont {Feng}}, \bibinfo
  {author} {\bibfnamefont {T.}~\bibnamefont {Onozato}}, \bibinfo {author}
  {\bibfnamefont {M.}~\bibnamefont {Wei}}, \bibinfo {author} {\bibfnamefont
  {A.~V.}\ \bibnamefont {Sanchela}}, \bibinfo {author} {\bibfnamefont
  {Y.}~\bibnamefont {Ikuhara}},\ and\ \bibinfo {author} {\bibfnamefont
  {H.}~\bibnamefont {Ohta}},\ }\href
  {https://doi.org/10.1103/PhysRevMaterials.3.094601} {\bibfield  {journal}
  {\bibinfo  {journal} {Phys. Rev. Mater.}\ }\textbf {\bibinfo {volume} {3}},\
  \bibinfo {pages} {094601} (\bibinfo {year} {2019}{\natexlab{a}})}\BibitemShut
  {NoStop}%
\bibitem [{\citenamefont {Postiglione}\ \emph {et~al.}(2021)\citenamefont
  {Postiglione}, \citenamefont {Ganguly}, \citenamefont {Yun}, \citenamefont
  {Jeong}, \citenamefont {Jacobson}, \citenamefont {Borgeson}, \citenamefont
  {Jalan}, \citenamefont {Mkhoyan},\ and\ \citenamefont
  {Leighton}}]{Postiglione2021}%
  \BibitemOpen
  \bibfield  {author} {\bibinfo {author} {\bibfnamefont {W.~M.}\ \bibnamefont
  {Postiglione}}, \bibinfo {author} {\bibfnamefont {K.}~\bibnamefont
  {Ganguly}}, \bibinfo {author} {\bibfnamefont {H.}~\bibnamefont {Yun}},
  \bibinfo {author} {\bibfnamefont {J.~S.}\ \bibnamefont {Jeong}}, \bibinfo
  {author} {\bibfnamefont {A.}~\bibnamefont {Jacobson}}, \bibinfo {author}
  {\bibfnamefont {L.}~\bibnamefont {Borgeson}}, \bibinfo {author}
  {\bibfnamefont {B.}~\bibnamefont {Jalan}}, \bibinfo {author} {\bibfnamefont
  {K.~A.}\ \bibnamefont {Mkhoyan}},\ and\ \bibinfo {author} {\bibfnamefont
  {C.}~\bibnamefont {Leighton}},\ }\href
  {https://doi.org/10.1103/PhysRevMaterials.5.044604} {\bibfield  {journal}
  {\bibinfo  {journal} {Phys. Rev. Mater.}\ }\textbf {\bibinfo {volume} {5}},\
  \bibinfo {pages} {044604} (\bibinfo {year} {2021})}\BibitemShut {NoStop}%
\bibitem [{\citenamefont {Sanchela}\ \emph {et~al.}(2018)\citenamefont
  {Sanchela}, \citenamefont {Wei}, \citenamefont {Zensyo}, \citenamefont
  {Feng}, \citenamefont {Lee}, \citenamefont {Kim}, \citenamefont {Jeen},
  \citenamefont {Ikuhara},\ and\ \citenamefont {Ohta}}]{Sanchela2018}%
  \BibitemOpen
  \bibfield  {author} {\bibinfo {author} {\bibfnamefont {A.~V.}\ \bibnamefont
  {Sanchela}}, \bibinfo {author} {\bibfnamefont {M.}~\bibnamefont {Wei}},
  \bibinfo {author} {\bibfnamefont {H.}~\bibnamefont {Zensyo}}, \bibinfo
  {author} {\bibfnamefont {B.}~\bibnamefont {Feng}}, \bibinfo {author}
  {\bibfnamefont {J.}~\bibnamefont {Lee}}, \bibinfo {author} {\bibfnamefont
  {G.}~\bibnamefont {Kim}}, \bibinfo {author} {\bibfnamefont {H.}~\bibnamefont
  {Jeen}}, \bibinfo {author} {\bibfnamefont {Y.}~\bibnamefont {Ikuhara}},\ and\
  \bibinfo {author} {\bibfnamefont {H.}~\bibnamefont {Ohta}},\ }\href
  {https://doi.org/10.1063/1.5033326} {\bibfield  {journal} {\bibinfo
  {journal} {Appl. Phys. Lett.}\ }\textbf {\bibinfo {volume} {112}},\ \bibinfo
  {pages} {232102} (\bibinfo {year} {2018})}\BibitemShut {NoStop}%
\bibitem [{\citenamefont {Prakash}\ \emph
  {et~al.}(2017{\natexlab{a}})\citenamefont {Prakash}, \citenamefont {Xu},
  \citenamefont {Faghaninia}, \citenamefont {Shukla}, \citenamefont {Ager},
  \citenamefont {Lo},\ and\ \citenamefont {Jalan}}]{Prakash2017c}%
  \BibitemOpen
  \bibfield  {author} {\bibinfo {author} {\bibfnamefont {A.}~\bibnamefont
  {Prakash}}, \bibinfo {author} {\bibfnamefont {P.}~\bibnamefont {Xu}},
  \bibinfo {author} {\bibfnamefont {A.}~\bibnamefont {Faghaninia}}, \bibinfo
  {author} {\bibfnamefont {S.}~\bibnamefont {Shukla}}, \bibinfo {author}
  {\bibfnamefont {J.~W.}\ \bibnamefont {Ager}}, \bibinfo {author}
  {\bibfnamefont {C.~S.}\ \bibnamefont {Lo}},\ and\ \bibinfo {author}
  {\bibfnamefont {B.}~\bibnamefont {Jalan}},\ }\href
  {https://doi.org/10.1038/ncomms15167} {\bibfield  {journal} {\bibinfo
  {journal} {Nat. Commun.}\ }\textbf {\bibinfo {volume} {8}},\ \bibinfo {pages}
  {15167} (\bibinfo {year} {2017}{\natexlab{a}})}\BibitemShut {NoStop}%
\bibitem [{\citenamefont {Yu}\ \emph {et~al.}(2016)\citenamefont {Yu},
  \citenamefont {Yoon},\ and\ \citenamefont {Son}}]{Yu2016}%
  \BibitemOpen
  \bibfield  {author} {\bibinfo {author} {\bibfnamefont {S.}~\bibnamefont
  {Yu}}, \bibinfo {author} {\bibfnamefont {D.}~\bibnamefont {Yoon}},\ and\
  \bibinfo {author} {\bibfnamefont {J.}~\bibnamefont {Son}},\ }\href
  {https://doi.org/10.1063/1.4954638} {\bibfield  {journal} {\bibinfo
  {journal} {Appl. Phys. Lett.}\ }\textbf {\bibinfo {volume} {108}},\ \bibinfo
  {pages} {262101} (\bibinfo {year} {2016})}\BibitemShut {NoStop}%
\bibitem [{\citenamefont {Cho}\ \emph {et~al.}(2019{\natexlab{b}})\citenamefont
  {Cho}, \citenamefont {Onozato}, \citenamefont {Wei}, \citenamefont
  {Sanchela},\ and\ \citenamefont {Ohta}}]{Cho2019}%
  \BibitemOpen
  \bibfield  {author} {\bibinfo {author} {\bibfnamefont {H.~J.}\ \bibnamefont
  {Cho}}, \bibinfo {author} {\bibfnamefont {T.}~\bibnamefont {Onozato}},
  \bibinfo {author} {\bibfnamefont {M.}~\bibnamefont {Wei}}, \bibinfo {author}
  {\bibfnamefont {A.}~\bibnamefont {Sanchela}},\ and\ \bibinfo {author}
  {\bibfnamefont {H.}~\bibnamefont {Ohta}},\ }\href
  {https://doi.org/10.1063/1.5054154} {\bibfield  {journal} {\bibinfo
  {journal} {APL Mater.}\ }\textbf {\bibinfo {volume} {7}},\ \bibinfo {pages}
  {022507} (\bibinfo {year} {2019}{\natexlab{b}})}\BibitemShut {NoStop}%
\bibitem [{\citenamefont {Paik}\ \emph {et~al.}(2017)\citenamefont {Paik},
  \citenamefont {Chen}, \citenamefont {Lochocki}, \citenamefont {Seidner~H.},
  \citenamefont {Verma}, \citenamefont {Tanen}, \citenamefont {Park},
  \citenamefont {Uchida}, \citenamefont {Shang}, \citenamefont {Zhou},
  \citenamefont {Brützam}, \citenamefont {Uecker}, \citenamefont {Liu},
  \citenamefont {Jena}, \citenamefont {Shen}, \citenamefont {Muller},\ and\
  \citenamefont {Schlom}}]{Paik2017c}%
  \BibitemOpen
  \bibfield  {author} {\bibinfo {author} {\bibfnamefont {H.}~\bibnamefont
  {Paik}}, \bibinfo {author} {\bibfnamefont {Z.}~\bibnamefont {Chen}}, \bibinfo
  {author} {\bibfnamefont {E.}~\bibnamefont {Lochocki}}, \bibinfo {author}
  {\bibfnamefont {A.}~\bibnamefont {Seidner~H.}}, \bibinfo {author}
  {\bibfnamefont {A.}~\bibnamefont {Verma}}, \bibinfo {author} {\bibfnamefont
  {N.}~\bibnamefont {Tanen}}, \bibinfo {author} {\bibfnamefont
  {J.}~\bibnamefont {Park}}, \bibinfo {author} {\bibfnamefont {M.}~\bibnamefont
  {Uchida}}, \bibinfo {author} {\bibfnamefont {S.}~\bibnamefont {Shang}},
  \bibinfo {author} {\bibfnamefont {B.-C.}\ \bibnamefont {Zhou}}, \bibinfo
  {author} {\bibfnamefont {M.}~\bibnamefont {Brützam}}, \bibinfo {author}
  {\bibfnamefont {R.}~\bibnamefont {Uecker}}, \bibinfo {author} {\bibfnamefont
  {Z.-K.}\ \bibnamefont {Liu}}, \bibinfo {author} {\bibfnamefont
  {D.}~\bibnamefont {Jena}}, \bibinfo {author} {\bibfnamefont {K.~M.}\
  \bibnamefont {Shen}}, \bibinfo {author} {\bibfnamefont {D.~A.}\ \bibnamefont
  {Muller}},\ and\ \bibinfo {author} {\bibfnamefont {D.~G.}\ \bibnamefont
  {Schlom}},\ }\href {https://doi.org/10.1063/1.5001839} {\bibfield  {journal}
  {\bibinfo  {journal} {APL Mater.}\ }\textbf {\bibinfo {volume} {5}},\
  \bibinfo {pages} {116107} (\bibinfo {year} {2017})}\BibitemShut {NoStop}%
\bibitem [{\citenamefont {Wadekar}\ \emph {et~al.}(2014)\citenamefont
  {Wadekar}, \citenamefont {Alaria}, \citenamefont {O'Sullivan}, \citenamefont
  {Flack}, \citenamefont {Manning}, \citenamefont {Phillips}, \citenamefont
  {Durose}, \citenamefont {Lozano}, \citenamefont {Lucas}, \citenamefont
  {Claridge},\ and\ \citenamefont {Rosseinsky}}]{PVWadekar_2014}%
  \BibitemOpen
  \bibfield  {author} {\bibinfo {author} {\bibfnamefont {P.~V.}\ \bibnamefont
  {Wadekar}}, \bibinfo {author} {\bibfnamefont {J.}~\bibnamefont {Alaria}},
  \bibinfo {author} {\bibfnamefont {M.}~\bibnamefont {O'Sullivan}}, \bibinfo
  {author} {\bibfnamefont {N.~L.~O.}\ \bibnamefont {Flack}}, \bibinfo {author}
  {\bibfnamefont {T.~D.}\ \bibnamefont {Manning}}, \bibinfo {author}
  {\bibfnamefont {L.~J.}\ \bibnamefont {Phillips}}, \bibinfo {author}
  {\bibfnamefont {K.}~\bibnamefont {Durose}}, \bibinfo {author} {\bibfnamefont
  {O.}~\bibnamefont {Lozano}}, \bibinfo {author} {\bibfnamefont
  {S.}~\bibnamefont {Lucas}}, \bibinfo {author} {\bibfnamefont {J.~B.}\
  \bibnamefont {Claridge}},\ and\ \bibinfo {author} {\bibfnamefont {M.~J.}\
  \bibnamefont {Rosseinsky}},\ }\href {https://doi.org/10.1063/1.4891816}
  {\bibfield  {journal} {\bibinfo  {journal} {Appl. Phys. Lett.}\ }\textbf
  {\bibinfo {volume} {105}},\ \bibinfo {pages} {052104} (\bibinfo {year}
  {2014})}\BibitemShut {NoStop}%
\bibitem [{\citenamefont {Nono~Tchiomo}\ \emph {et~al.}(2019)\citenamefont
  {Nono~Tchiomo}, \citenamefont {Braun}, \citenamefont {Doyle}, \citenamefont
  {Sigle}, \citenamefont {van Aken}, \citenamefont {Mannhart},\ and\
  \citenamefont {Ngabonziza}}]{Nono-Tchiomo2019}%
  \BibitemOpen
  \bibfield  {author} {\bibinfo {author} {\bibfnamefont {A.~P.}\ \bibnamefont
  {Nono~Tchiomo}}, \bibinfo {author} {\bibfnamefont {W.}~\bibnamefont {Braun}},
  \bibinfo {author} {\bibfnamefont {B.~P.}\ \bibnamefont {Doyle}}, \bibinfo
  {author} {\bibfnamefont {W.}~\bibnamefont {Sigle}}, \bibinfo {author}
  {\bibfnamefont {P.}~\bibnamefont {van Aken}}, \bibinfo {author}
  {\bibfnamefont {J.}~\bibnamefont {Mannhart}},\ and\ \bibinfo {author}
  {\bibfnamefont {P.}~\bibnamefont {Ngabonziza}},\ }\href
  {https://doi.org/10.1063/1.5094867} {\bibfield  {journal} {\bibinfo
  {journal} {APL Mater.}\ }\textbf {\bibinfo {volume} {7}},\ \bibinfo {pages}
  {041119} (\bibinfo {year} {2019})}\BibitemShut {NoStop}%
\bibitem [{\citenamefont {He}\ \emph {et~al.}(2021)\citenamefont {He},
  \citenamefont {Wei}, \citenamefont {Zhou}, \citenamefont {Cheng},
  \citenamefont {Ding}, \citenamefont {Shao}, \citenamefont {Hu}, \citenamefont
  {Song}, \citenamefont {Zhu},\ and\ \citenamefont {Sun}}]{YHe_2021}%
  \BibitemOpen
  \bibfield  {author} {\bibinfo {author} {\bibfnamefont {Y.}~\bibnamefont
  {He}}, \bibinfo {author} {\bibfnamefont {R.}~\bibnamefont {Wei}}, \bibinfo
  {author} {\bibfnamefont {C.}~\bibnamefont {Zhou}}, \bibinfo {author}
  {\bibfnamefont {W.}~\bibnamefont {Cheng}}, \bibinfo {author} {\bibfnamefont
  {X.}~\bibnamefont {Ding}}, \bibinfo {author} {\bibfnamefont {C.}~\bibnamefont
  {Shao}}, \bibinfo {author} {\bibfnamefont {L.}~\bibnamefont {Hu}}, \bibinfo
  {author} {\bibfnamefont {W.}~\bibnamefont {Song}}, \bibinfo {author}
  {\bibfnamefont {X.}~\bibnamefont {Zhu}},\ and\ \bibinfo {author}
  {\bibfnamefont {Y.}~\bibnamefont {Sun}},\ }\href
  {https://doi.org/10.1021/acs.cgd.1c00698} {\bibfield  {journal} {\bibinfo
  {journal} {Cryst. Growth Des.}\ }\textbf {\bibinfo {volume} {21}},\ \bibinfo
  {pages} {5800} (\bibinfo {year} {2021})}\BibitemShut {NoStop}%
\bibitem [{\citenamefont {Lebens-Higgins}\ \emph {et~al.}(2016)\citenamefont
  {Lebens-Higgins}, \citenamefont {Scanlon}, \citenamefont {Paik},
  \citenamefont {Sallis}, \citenamefont {Nie}, \citenamefont {Uchida},
  \citenamefont {Quackenbush}, \citenamefont {Wahila}, \citenamefont
  {Sterbinsky}, \citenamefont {Arena}, \citenamefont {Woicik}, \citenamefont
  {Schlom},\ and\ \citenamefont {Piper}}]{Lebens-Higgins2016a}%
  \BibitemOpen
  \bibfield  {author} {\bibinfo {author} {\bibfnamefont {Z.}~\bibnamefont
  {Lebens-Higgins}}, \bibinfo {author} {\bibfnamefont {D.~O.}\ \bibnamefont
  {Scanlon}}, \bibinfo {author} {\bibfnamefont {H.}~\bibnamefont {Paik}},
  \bibinfo {author} {\bibfnamefont {S.}~\bibnamefont {Sallis}}, \bibinfo
  {author} {\bibfnamefont {Y.}~\bibnamefont {Nie}}, \bibinfo {author}
  {\bibfnamefont {M.}~\bibnamefont {Uchida}}, \bibinfo {author} {\bibfnamefont
  {N.~F.}\ \bibnamefont {Quackenbush}}, \bibinfo {author} {\bibfnamefont
  {M.~J.}\ \bibnamefont {Wahila}}, \bibinfo {author} {\bibfnamefont {G.~E.}\
  \bibnamefont {Sterbinsky}}, \bibinfo {author} {\bibfnamefont {D.~A.}\
  \bibnamefont {Arena}}, \bibinfo {author} {\bibfnamefont {J.~C.}\ \bibnamefont
  {Woicik}}, \bibinfo {author} {\bibfnamefont {D.~G.}\ \bibnamefont {Schlom}},\
  and\ \bibinfo {author} {\bibfnamefont {L.~F.~J.}\ \bibnamefont {Piper}},\
  }\href {https://doi.org/10.1103/PhysRevLett.116.027602} {\bibfield  {journal}
  {\bibinfo  {journal} {Phys. Rev. Lett.}\ }\textbf {\bibinfo {volume} {116}},\
  \bibinfo {pages} {027602} (\bibinfo {year} {2016})}\BibitemShut {NoStop}%
\bibitem [{\citenamefont {Zhang}\ \emph {et~al.}(2021)\citenamefont {Zhang},
  \citenamefont {Li}, \citenamefont {Bi}, \citenamefont {Zhang}, \citenamefont
  {Peng}, \citenamefont {Song}, \citenamefont {Zhang}, \citenamefont {Gu},
  \citenamefont {Duan},\ and\ \citenamefont {Cao}}]{RZhang_2021}%
  \BibitemOpen
  \bibfield  {author} {\bibinfo {author} {\bibfnamefont {R.}~\bibnamefont
  {Zhang}}, \bibinfo {author} {\bibfnamefont {X.}~\bibnamefont {Li}}, \bibinfo
  {author} {\bibfnamefont {J.}~\bibnamefont {Bi}}, \bibinfo {author}
  {\bibfnamefont {S.}~\bibnamefont {Zhang}}, \bibinfo {author} {\bibfnamefont
  {S.}~\bibnamefont {Peng}}, \bibinfo {author} {\bibfnamefont {Y.}~\bibnamefont
  {Song}}, \bibinfo {author} {\bibfnamefont {Q.}~\bibnamefont {Zhang}},
  \bibinfo {author} {\bibfnamefont {L.}~\bibnamefont {Gu}}, \bibinfo {author}
  {\bibfnamefont {J.}~\bibnamefont {Duan}},\ and\ \bibinfo {author}
  {\bibfnamefont {Y.}~\bibnamefont {Cao}},\ }\href
  {https://doi.org/10.1063/5.0046639} {\bibfield  {journal} {\bibinfo
  {journal} {APL Mater.}\ }\textbf {\bibinfo {volume} {9}},\ \bibinfo {pages}
  {061103} (\bibinfo {year} {2021})}\BibitemShut {NoStop}%
\bibitem [{\citenamefont {Murauskas}\ \emph {et~al.}(2022)\citenamefont
  {Murauskas}, \citenamefont {Kubilius}, \citenamefont {Talaikis},
  \citenamefont {Abrutis}, \citenamefont {Raudonis}, \citenamefont {Niaura},\
  and\ \citenamefont {Plausinaitiene}}]{TMurauskas_2022}%
  \BibitemOpen
  \bibfield  {author} {\bibinfo {author} {\bibfnamefont {T.}~\bibnamefont
  {Murauskas}}, \bibinfo {author} {\bibfnamefont {V.}~\bibnamefont {Kubilius}},
  \bibinfo {author} {\bibfnamefont {M.}~\bibnamefont {Talaikis}}, \bibinfo
  {author} {\bibfnamefont {A.}~\bibnamefont {Abrutis}}, \bibinfo {author}
  {\bibfnamefont {R.}~\bibnamefont {Raudonis}}, \bibinfo {author}
  {\bibfnamefont {G.}~\bibnamefont {Niaura}},\ and\ \bibinfo {author}
  {\bibfnamefont {V.}~\bibnamefont {Plausinaitiene}},\ }\href
  {https://doi.org/https://doi.org/10.1016/j.jallcom.2021.162843} {\bibfield
  {journal} {\bibinfo  {journal} {J. Alloys Compd.}\ }\textbf {\bibinfo
  {volume} {898}},\ \bibinfo {pages} {162843} (\bibinfo {year}
  {2022})}\BibitemShut {NoStop}%
\bibitem [{\citenamefont {Shiogai}\ \emph {et~al.}(2016)\citenamefont
  {Shiogai}, \citenamefont {Nishihara}, \citenamefont {Sato},\ and\
  \citenamefont {Tsukazaki}}]{JShiogai_2016}%
  \BibitemOpen
  \bibfield  {author} {\bibinfo {author} {\bibfnamefont {J.}~\bibnamefont
  {Shiogai}}, \bibinfo {author} {\bibfnamefont {K.}~\bibnamefont {Nishihara}},
  \bibinfo {author} {\bibfnamefont {K.}~\bibnamefont {Sato}},\ and\ \bibinfo
  {author} {\bibfnamefont {A.}~\bibnamefont {Tsukazaki}},\ }\href
  {https://doi.org/10.1063/1.4953808} {\bibfield  {journal} {\bibinfo
  {journal} {AIP Advances}\ }\textbf {\bibinfo {volume} {6}},\ \bibinfo {pages}
  {065305} (\bibinfo {year} {2016})}\BibitemShut {NoStop}%
\bibitem [{\citenamefont {Eom}\ \emph {et~al.}(2022)\citenamefont {Eom},
  \citenamefont {Paik}, \citenamefont {Seo}, \citenamefont {Campbell},
  \citenamefont {Tsymbal}, \citenamefont {Oh}, \citenamefont {Rzchowski},
  \citenamefont {Schlom},\ and\ \citenamefont {Eom}}]{KEom_2021}%
  \BibitemOpen
  \bibfield  {author} {\bibinfo {author} {\bibfnamefont {K.}~\bibnamefont
  {Eom}}, \bibinfo {author} {\bibfnamefont {H.}~\bibnamefont {Paik}}, \bibinfo
  {author} {\bibfnamefont {J.}~\bibnamefont {Seo}}, \bibinfo {author}
  {\bibfnamefont {N.}~\bibnamefont {Campbell}}, \bibinfo {author}
  {\bibfnamefont {E.~Y.}\ \bibnamefont {Tsymbal}}, \bibinfo {author}
  {\bibfnamefont {S.~H.}\ \bibnamefont {Oh}}, \bibinfo {author} {\bibfnamefont
  {M.~S.}\ \bibnamefont {Rzchowski}}, \bibinfo {author} {\bibfnamefont {D.~G.}\
  \bibnamefont {Schlom}},\ and\ \bibinfo {author} {\bibfnamefont {C.-B.}\
  \bibnamefont {Eom}},\ }\href
  {https://doi.org/https://doi.org/10.1002/advs.202105652} {\bibfield
  {journal} {\bibinfo  {journal} {Adv. Sci.}\ }\textbf {\bibinfo {volume}
  {9}},\ \bibinfo {pages} {2105652} (\bibinfo {year} {2022})}\BibitemShut
  {NoStop}%
\bibitem [{\citenamefont {Kang}\ \emph {et~al.}(2022)\citenamefont {Kang},
  \citenamefont {Lee}, \citenamefont {Lee}, \citenamefont {Kim}, \citenamefont
  {Kim}, \citenamefont {Maeng}, \citenamefont {Hong}, \citenamefont {Park},\
  and\ \citenamefont {Kim}}]{JKang2022}%
  \BibitemOpen
  \bibfield  {author} {\bibinfo {author} {\bibfnamefont {J.}~\bibnamefont
  {Kang}}, \bibinfo {author} {\bibfnamefont {J.~H.}\ \bibnamefont {Lee}},
  \bibinfo {author} {\bibfnamefont {H.-K.}\ \bibnamefont {Lee}}, \bibinfo
  {author} {\bibfnamefont {K.-T.}\ \bibnamefont {Kim}}, \bibinfo {author}
  {\bibfnamefont {J.~H.}\ \bibnamefont {Kim}}, \bibinfo {author} {\bibfnamefont
  {M.-J.}\ \bibnamefont {Maeng}}, \bibinfo {author} {\bibfnamefont {J.-A.}\
  \bibnamefont {Hong}}, \bibinfo {author} {\bibfnamefont {Y.}~\bibnamefont
  {Park}},\ and\ \bibinfo {author} {\bibfnamefont {K.~H.}\ \bibnamefont
  {Kim}},\ }\bibfield  {journal} {\bibinfo  {journal} {Materials}\ }\textbf
  {\bibinfo {volume} {15}},\ \href {https://doi.org/10.3390/ma15072417}
  {10.3390/ma15072417} (\bibinfo {year} {2022})\BibitemShut {NoStop}%
\bibitem [{\citenamefont {Gesing}\ \emph {et~al.}(2009)\citenamefont {Gesing},
  \citenamefont {Uecker},\ and\ \citenamefont {Buhl}}]{TMGesing2009}%
  \BibitemOpen
  \bibfield  {author} {\bibinfo {author} {\bibfnamefont {T.~M.}\ \bibnamefont
  {Gesing}}, \bibinfo {author} {\bibfnamefont {R.}~\bibnamefont {Uecker}},\
  and\ \bibinfo {author} {\bibfnamefont {J.-C.}\ \bibnamefont {Buhl}},\ }\href
  {https://doi.org/doi:10.1524/ncrs.2009.0159} {\bibfield  {journal} {\bibinfo
  {journal} {Kristallogr. NCS}\ }\textbf {\bibinfo {volume} {224}},\ \bibinfo
  {pages} {365} (\bibinfo {year} {2009})}\BibitemShut {NoStop}%
\bibitem [{\citenamefont {Wang}\ \emph {et~al.}(2015)\citenamefont {Wang},
  \citenamefont {Tang}, \citenamefont {Zhu}, \citenamefont {Suriyaprakash},
  \citenamefont {Xu}, \citenamefont {Liu}, \citenamefont {Gao}, \citenamefont
  {Cheong},\ and\ \citenamefont {Ma}}]{WYWang2015}%
  \BibitemOpen
  \bibfield  {author} {\bibinfo {author} {\bibfnamefont {W.~Y.}\ \bibnamefont
  {Wang}}, \bibinfo {author} {\bibfnamefont {Y.~L.}\ \bibnamefont {Tang}},
  \bibinfo {author} {\bibfnamefont {Y.~L.}\ \bibnamefont {Zhu}}, \bibinfo
  {author} {\bibfnamefont {J.}~\bibnamefont {Suriyaprakash}}, \bibinfo {author}
  {\bibfnamefont {Y.~B.}\ \bibnamefont {Xu}}, \bibinfo {author} {\bibfnamefont
  {Y.}~\bibnamefont {Liu}}, \bibinfo {author} {\bibfnamefont {B.}~\bibnamefont
  {Gao}}, \bibinfo {author} {\bibfnamefont {S.-W.}\ \bibnamefont {Cheong}},\
  and\ \bibinfo {author} {\bibfnamefont {X.~L.}\ \bibnamefont {Ma}},\ }\href
  {https://doi.org/10.1038/srep16097} {\bibfield  {journal} {\bibinfo
  {journal} {Sci Rep}\ }\textbf {\bibinfo {volume} {5}},\ \bibinfo {pages}
  {16097} (\bibinfo {year} {2015})}\BibitemShut {NoStop}%
\bibitem [{\citenamefont {Raghavan}\ \emph {et~al.}(2016)\citenamefont
  {Raghavan}, \citenamefont {Schumann}, \citenamefont {Kim}, \citenamefont
  {Zhang}, \citenamefont {Cain},\ and\ \citenamefont
  {Stemmer}}]{Raghavan2016c}%
  \BibitemOpen
  \bibfield  {author} {\bibinfo {author} {\bibfnamefont {S.}~\bibnamefont
  {Raghavan}}, \bibinfo {author} {\bibfnamefont {T.}~\bibnamefont {Schumann}},
  \bibinfo {author} {\bibfnamefont {H.}~\bibnamefont {Kim}}, \bibinfo {author}
  {\bibfnamefont {J.~Y.}\ \bibnamefont {Zhang}}, \bibinfo {author}
  {\bibfnamefont {T.~A.}\ \bibnamefont {Cain}},\ and\ \bibinfo {author}
  {\bibfnamefont {S.}~\bibnamefont {Stemmer}},\ }\href
  {https://doi.org/10.1063/1.4939657} {\bibfield  {journal} {\bibinfo
  {journal} {APL Mater.}\ }\textbf {\bibinfo {volume} {4}},\ \bibinfo {pages}
  {016106} (\bibinfo {year} {2016})}\BibitemShut {NoStop}%
\bibitem [{\citenamefont {Shin}\ \emph {et~al.}(2016)\citenamefont {Shin},
  \citenamefont {Kim}, \citenamefont {Kim}, \citenamefont {Park},\ and\
  \citenamefont {Char}}]{SJuyeon_2016}%
  \BibitemOpen
  \bibfield  {author} {\bibinfo {author} {\bibfnamefont {J.}~\bibnamefont
  {Shin}}, \bibinfo {author} {\bibfnamefont {Y.~M.}\ \bibnamefont {Kim}},
  \bibinfo {author} {\bibfnamefont {Y.}~\bibnamefont {Kim}}, \bibinfo {author}
  {\bibfnamefont {C.}~\bibnamefont {Park}},\ and\ \bibinfo {author}
  {\bibfnamefont {K.}~\bibnamefont {Char}},\ }\href
  {https://doi.org/10.1063/1.4973205} {\bibfield  {journal} {\bibinfo
  {journal} {Appl. Phys. Lett.}\ }\textbf {\bibinfo {volume} {109}},\ \bibinfo
  {pages} {262102} (\bibinfo {year} {2016})}\BibitemShut {NoStop}%
\bibitem [{\citenamefont {Fujiwara}\ \emph {et~al.}(2016)\citenamefont
  {Fujiwara}, \citenamefont {Nishihara}, \citenamefont {Shiogai},\ and\
  \citenamefont {Tsukazaki}}]{FKohei_2016}%
  \BibitemOpen
  \bibfield  {author} {\bibinfo {author} {\bibfnamefont {K.}~\bibnamefont
  {Fujiwara}}, \bibinfo {author} {\bibfnamefont {K.}~\bibnamefont {Nishihara}},
  \bibinfo {author} {\bibfnamefont {J.}~\bibnamefont {Shiogai}},\ and\ \bibinfo
  {author} {\bibfnamefont {A.}~\bibnamefont {Tsukazaki}},\ }\href
  {https://doi.org/10.1063/1.4961637} {\bibfield  {journal} {\bibinfo
  {journal} {AIP Advances}\ }\textbf {\bibinfo {volume} {6}},\ \bibinfo {pages}
  {085014} (\bibinfo {year} {2016})}\BibitemShut {NoStop}%
\bibitem [{\citenamefont {Lee}\ \emph {et~al.}(2016)\citenamefont {Lee},
  \citenamefont {Kim}, \citenamefont {Sohn}, \citenamefont {Kim}, \citenamefont
  {Park}, \citenamefont {Park}, \citenamefont {Jeong}, \citenamefont {Lee},
  \citenamefont {Kim}, \citenamefont {Choi},\ and\ \citenamefont
  {Kim}}]{LWoong-Jhae_2016}%
  \BibitemOpen
  \bibfield  {author} {\bibinfo {author} {\bibfnamefont {W.-J.}\ \bibnamefont
  {Lee}}, \bibinfo {author} {\bibfnamefont {H.~J.}\ \bibnamefont {Kim}},
  \bibinfo {author} {\bibfnamefont {E.}~\bibnamefont {Sohn}}, \bibinfo {author}
  {\bibfnamefont {T.~H.}\ \bibnamefont {Kim}}, \bibinfo {author} {\bibfnamefont
  {J.-Y.}\ \bibnamefont {Park}}, \bibinfo {author} {\bibfnamefont
  {W.}~\bibnamefont {Park}}, \bibinfo {author} {\bibfnamefont {H.}~\bibnamefont
  {Jeong}}, \bibinfo {author} {\bibfnamefont {T.}~\bibnamefont {Lee}}, \bibinfo
  {author} {\bibfnamefont {J.~H.}\ \bibnamefont {Kim}}, \bibinfo {author}
  {\bibfnamefont {K.-Y.}\ \bibnamefont {Choi}},\ and\ \bibinfo {author}
  {\bibfnamefont {K.~H.}\ \bibnamefont {Kim}},\ }\href
  {https://doi.org/10.1063/1.4942509} {\bibfield  {journal} {\bibinfo
  {journal} {Appl. Phys. Lett.}\ }\textbf {\bibinfo {volume} {108}},\ \bibinfo
  {pages} {082105} (\bibinfo {year} {2016})}\BibitemShut {NoStop}%
\bibitem [{\citenamefont {Yoon}\ \emph {et~al.}(2018)\citenamefont {Yoon},
  \citenamefont {Yu},\ and\ \citenamefont {Son}}]{Yoon2018}%
  \BibitemOpen
  \bibfield  {author} {\bibinfo {author} {\bibfnamefont {D.}~\bibnamefont
  {Yoon}}, \bibinfo {author} {\bibfnamefont {S.}~\bibnamefont {Yu}},\ and\
  \bibinfo {author} {\bibfnamefont {J.}~\bibnamefont {Son}},\ }\href
  {https://doi.org/10.1038/s41427-018-0038-1} {\bibfield  {journal} {\bibinfo
  {journal} {NPG Asia Mater.}\ }\textbf {\bibinfo {volume} {10}},\ \bibinfo
  {pages} {363} (\bibinfo {year} {2018})}\BibitemShut {NoStop}%
\bibitem [{\citenamefont {Prakash}\ \emph
  {et~al.}(2017{\natexlab{b}})\citenamefont {Prakash}, \citenamefont {Xu},
  \citenamefont {Wu}, \citenamefont {Haugstad}, \citenamefont {Wang},\ and\
  \citenamefont {Jalan}}]{APrakash_2017}%
  \BibitemOpen
  \bibfield  {author} {\bibinfo {author} {\bibfnamefont {A.}~\bibnamefont
  {Prakash}}, \bibinfo {author} {\bibfnamefont {P.}~\bibnamefont {Xu}},
  \bibinfo {author} {\bibfnamefont {X.}~\bibnamefont {Wu}}, \bibinfo {author}
  {\bibfnamefont {G.}~\bibnamefont {Haugstad}}, \bibinfo {author}
  {\bibfnamefont {X.}~\bibnamefont {Wang}},\ and\ \bibinfo {author}
  {\bibfnamefont {B.}~\bibnamefont {Jalan}},\ }\href
  {https://doi.org/10.1039/C7TC00190H} {\bibfield  {journal} {\bibinfo
  {journal} {J. Mater. Chem. C}\ }\textbf {\bibinfo {volume} {5}},\ \bibinfo
  {pages} {5730} (\bibinfo {year} {2017}{\natexlab{b}})}\BibitemShut {NoStop}%
\bibitem [{\citenamefont {Braun}\ \emph {et~al.}(2020)\citenamefont {Braun},
  \citenamefont {Jäger}, \citenamefont {Laskin}, \citenamefont {Ngabonziza},
  \citenamefont {Voesch}, \citenamefont {Wittlich},\ and\ \citenamefont
  {Mannhart}}]{WBraun_2020}%
  \BibitemOpen
  \bibfield  {author} {\bibinfo {author} {\bibfnamefont {W.}~\bibnamefont
  {Braun}}, \bibinfo {author} {\bibfnamefont {M.}~\bibnamefont {Jäger}},
  \bibinfo {author} {\bibfnamefont {G.}~\bibnamefont {Laskin}}, \bibinfo
  {author} {\bibfnamefont {P.}~\bibnamefont {Ngabonziza}}, \bibinfo {author}
  {\bibfnamefont {W.}~\bibnamefont {Voesch}}, \bibinfo {author} {\bibfnamefont
  {P.}~\bibnamefont {Wittlich}},\ and\ \bibinfo {author} {\bibfnamefont
  {J.}~\bibnamefont {Mannhart}},\ }\href {https://doi.org/10.1063/5.0008324}
  {\bibfield  {journal} {\bibinfo  {journal} {APL Mater.}\ }\textbf {\bibinfo
  {volume} {8}},\ \bibinfo {pages} {071112} (\bibinfo {year}
  {2020})}\BibitemShut {NoStop}%
\bibitem [{\citenamefont {Luo}\ \emph {et~al.}(2015)\citenamefont {Luo},
  \citenamefont {Riggs}, \citenamefont {Zhang}, \citenamefont {Shipman},
  \citenamefont {Adireddy}, \citenamefont {Sklare}, \citenamefont {Koplitz},\
  and\ \citenamefont {Chrisey}}]{LSijun_2015}%
  \BibitemOpen
  \bibfield  {author} {\bibinfo {author} {\bibfnamefont {S.}~\bibnamefont
  {Luo}}, \bibinfo {author} {\bibfnamefont {B.~C.}\ \bibnamefont {Riggs}},
  \bibinfo {author} {\bibfnamefont {X.}~\bibnamefont {Zhang}}, \bibinfo
  {author} {\bibfnamefont {J.~T.}\ \bibnamefont {Shipman}}, \bibinfo {author}
  {\bibfnamefont {S.}~\bibnamefont {Adireddy}}, \bibinfo {author}
  {\bibfnamefont {S.~C.}\ \bibnamefont {Sklare}}, \bibinfo {author}
  {\bibfnamefont {B.}~\bibnamefont {Koplitz}},\ and\ \bibinfo {author}
  {\bibfnamefont {D.~B.}\ \bibnamefont {Chrisey}},\ }\href
  {https://doi.org/10.1063/1.4927158} {\bibfield  {journal} {\bibinfo
  {journal} {J. Appl. Phys.}\ }\textbf {\bibinfo {volume} {118}},\ \bibinfo
  {pages} {035310} (\bibinfo {year} {2015})}\BibitemShut {NoStop}%
\bibitem [{Not()}]{Note_01}%
  \BibitemOpen
  \href@noop {} {\bibinfo  {journal} {After their epitaxial growth by PLD, the
  high temperature-grown SrZrO$_3$ layers were shipped under ambient conditions
  for the subsequent MBE growth of the overlying La:BaSnO$_3$ active layers.
  Despite an extended exposure of SrZrO$_3$ buffer layers to ambient
  conditions, the extracted density of TDs is still very low, thus highlighting
  the stability and robustness of the high-temperature-grown SrZrO$_3$ buffer
  layers as a suitable template for subsequent growth of low defect density,
  high mobility La:BaSnO$_3$ films}\ }\BibitemShut {NoStop}%
\bibitem [{\citenamefont {Mun}\ \emph {et~al.}(2013)\citenamefont {Mun},
  \citenamefont {Kim}, \citenamefont {Min~Kim}, \citenamefont {Park},
  \citenamefont {Hoon~Kim}, \citenamefont {Joon~Kim}, \citenamefont
  {Hoon~Kim},\ and\ \citenamefont {Char}}]{HMun_2013}%
  \BibitemOpen
\bibfield  {journal} {  }\bibfield  {author} {\bibinfo {author} {\bibfnamefont
  {H.}~\bibnamefont {Mun}}, \bibinfo {author} {\bibfnamefont {U.}~\bibnamefont
  {Kim}}, \bibinfo {author} {\bibfnamefont {H.}~\bibnamefont {Min~Kim}},
  \bibinfo {author} {\bibfnamefont {C.}~\bibnamefont {Park}}, \bibinfo {author}
  {\bibfnamefont {T.}~\bibnamefont {Hoon~Kim}}, \bibinfo {author}
  {\bibfnamefont {H.}~\bibnamefont {Joon~Kim}}, \bibinfo {author}
  {\bibfnamefont {K.}~\bibnamefont {Hoon~Kim}},\ and\ \bibinfo {author}
  {\bibfnamefont {K.}~\bibnamefont {Char}},\ }\href
  {https://doi.org/10.1063/1.4812642} {\bibfield  {journal} {\bibinfo
  {journal} {Appl. Phys. Lett.}\ }\textbf {\bibinfo {volume} {102}},\ \bibinfo
  {pages} {252105} (\bibinfo {year} {2013})}\BibitemShut {NoStop}%
\end{thebibliography}%

\onecolumngrid
\newpage
\setcounter{table}{0}
\setcounter{figure}{0}
\renewcommand{\thefigure}{S\arabic{figure}}%
\setcounter{equation}{0}
\renewcommand{\theequation}{S\arabic{equation}}%
\setstretch{1.5}
\begin{center}
\title*{\textbf{\large{Supplementary Information:} \\ [0.25in] \large{Employing High-temperature-grown SrZrO$_3$ Buffer to Enhance the Electron Mobility in La:BaSnO$_3$-based Heterostructures}}}
\end{center}
\begin{center}
\large{Prosper Ngabonziza,$^{1,2,{\textcolor{blue}{\small{*}}}}$, Jisung Park,$^3$  Wilfried Sigle,$^4$ \\ \; \;\;\;\;\;\;\;\;\;\;\;\;\;\;\;Peter A. van Aken,$^4$ Jochen Mannhart,$^4$  and Darrell G. Schlom,$^{3,\,5,\,6}$}\newline 
\\
\large{\textit{$^1$Department of Physics $\&$ Astronomy , Louisiana State University, Baton Rouge, LA 70803, USA
\newline $^2$Department of Physics, University of Johannesburg, P.O. Box 524,  Auckland Park 2006,  Johannesburg, South Africa \newline $^3$Department of Material Science $\&$ Engineering, Cornell University, Ithaca, New York 14853, USA \newline  $^4$ Max Planck Institute for Solid State Research, Heisenbergstr. 1, 70569 Stuttgart, Germany \newline $^5$Kavli Institute at Cornell for Nanoscale Science, Ithaca, New York 14853, USA \newline $^6$ Leibniz-Institut für Kristallzüchtung, Max-Born-Str. 2, 12489 Berlin, Germany }}
\end{center}

\begin{center}
\textbf{\large{RHEED Intensity Oscillations of the PLD-grown SrZrO$_3$ Buffer Layers}}
\end{center}

\begin{figure*}[!b]
     \includegraphics[width=1\textwidth]{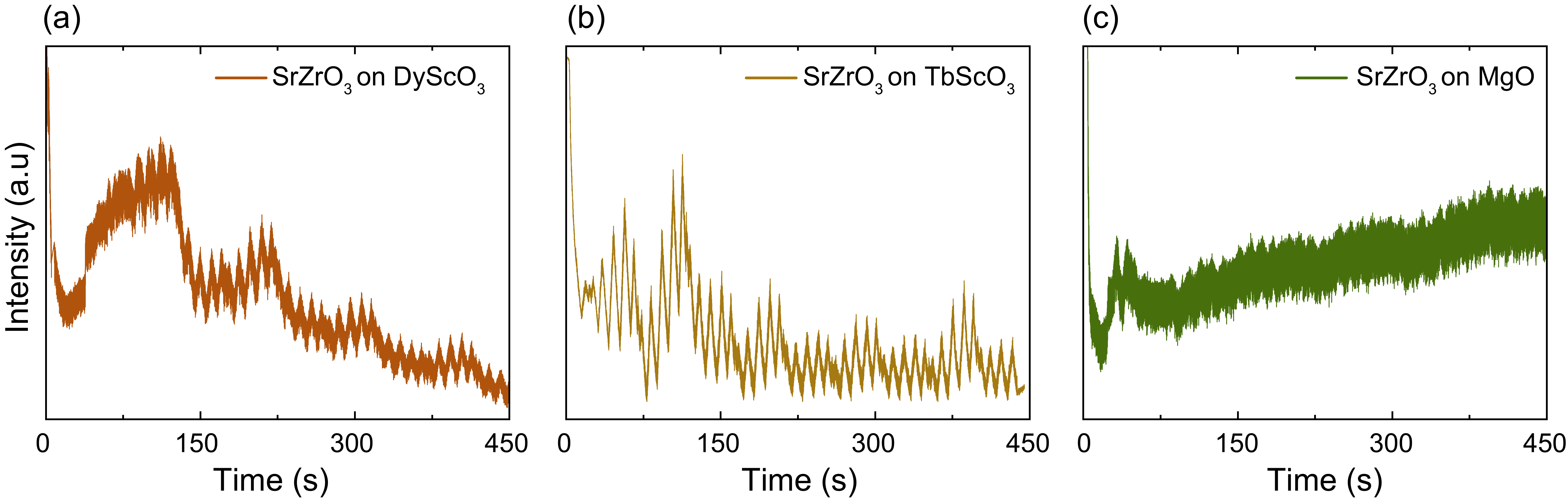}
    \caption{\small{RHEED intensity oscillations during the PLD growth of SrZrO$_3$ buffer layers grown on (a) DyScO$_3$ (b)  TbScO$_3$, and (c) MgO substrates.}} 
\label{Fig_S01}
\end{figure*}

For the pulsed laser deposition (PLD) growth of the SrZrO$_3$ buffer layers, we acquired reflection high-energy electron diffraction (RHEED) images and oscillations. The RHEED gun was a differentially pumped by Staib system operated at 30 keV. We show below RHEED data of the PLD-grown SrZrO$_3$ buffer layers deposited on $(110)$-oriented DyScO$_3$ [Fig.~\ref{Fig_S01}\textcolor{blue}{(a)}], (b)  $(110)$-oriented TbScO$_3$ [Fig.~\ref{Fig_S01}\textcolor{blue}{(b)}], and (c) $(001)$-oriented MgO [Fig.~\ref{Fig_S01}\textcolor{blue}{(c)}] single crystalline substrates. The observed RHEED oscillations are consistent  with those previously reported for high-temperature PLD-grown SrZrO$_3$ buffer layers~\cite{Nono-Tchiomo2019}.
\newpage
\begin{center}
\textbf{\large{Additional  Microstructural Characterizations of the La:BaSnO$_3$/SrZrO$_3$ Heterostructures}}
\end{center}

\begin{figure*}[!h]
     \includegraphics[width=1\textwidth]{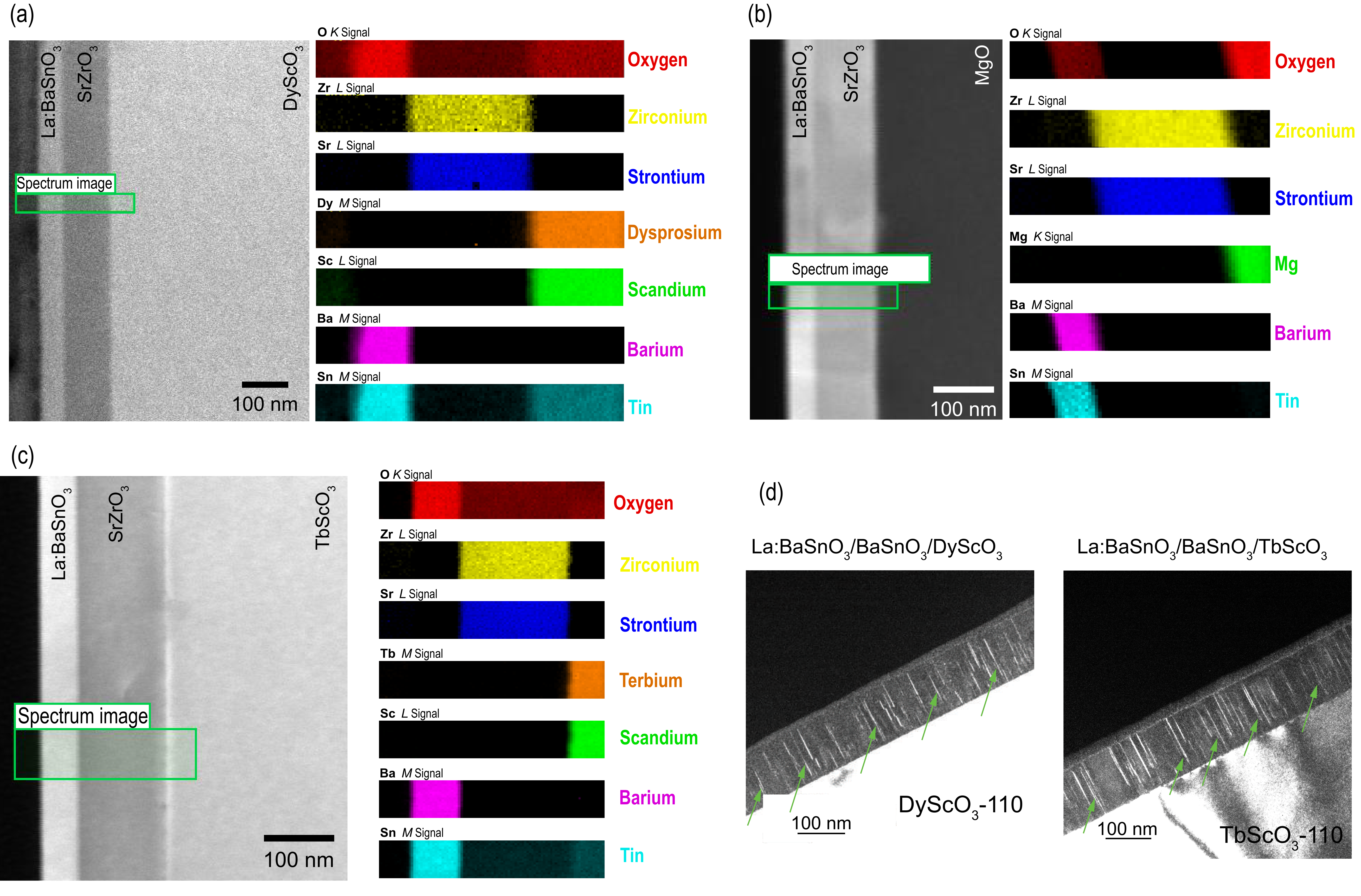}
    \caption{\small{Large-scale TEM images and corresponding EELS  elemental  maps of representative La:BaSnO$_3$/SrZrO$_3$ heterostructure grown on (a) $(110)$-oriented DyScO$_3$,(b) $(001)$-oriented MgO, and (c) $(110)$-oriented TbScO$_3$   single crystalline substrates. There are no discernible edge-type TDs in these TEM images. (d) In contrast, for similar La:BaSnO$_3$-BaSnO$_3$ films with BaSnO$_3$ grown at 850\degree C, TDs are observed (vertical bright
contrasts) running across the film from the interface, as indicated by green arrows (Fig.~\ref{Fig_S02}\textcolor{blue}{(d)} is adapted from Ref.~[\textcolor{blue}{2}]).}} 
\label{Fig_S02}
\end{figure*}

The scanning transmission electron microscopy (STEM) investigations were performed using a Cs-probe-corrected JEOL JEM-ARM200F. We present here additional STEM and electron  energy-loss  spectroscopy (EELS) data over a wider region for the La:BaSnO$_3$/SrZrO$_3$ heterostructures [Fig.~\ref{Fig_S02}\textcolor{blue}{(a)}-\textcolor{blue}{(c)}]. For areas  analyzed in these TEM data, EELS elemental maps show expected composition in the films. Also, no edge-type threading dislocations (TDs) are observed in the bright-field TEM images of the entire film thickness, which highlight the efficacy of inserting a high-temperature-grown SrZrO$_3$ layer between the La:BaSnO$_3$ film and substrates. The primary role of this high-temperature-grown buffer is to minimize the density of defects. In particular, the 
concentration of TDs  is dramatically decreased, which results in carrier density and carrier mobility improvement~[\textcolor{blue}{1}].

	Although both XRD and TEM analysis indicate significant mismatch at the interfaces between SrZrO$_3$ films and the substrates, there were no noticeable apparent structural defects and stacking faults across the entire La:BaSnO$_3$/SrZrO$_3$ heterostructure running from the interface to substrates [Fig.~\ref{Fig_05} and Fig.~\ref{Fig_S02}\textcolor{blue}{(a)}-\textcolor{blue}{(c)}]. Further, the dislocation density in the La:BaSnO$_3$ active layers is very low resulting in high RT  $\mu_e$. In contrast, analysis of control samples of  La:BaSnO$_3$/BaSnO$_3$ heterostructures with the BaSnO$_3$ buffer layers grown at relatively low temperature (850\degree C), their TEM micrographs show higher density of edge-type TDs, as identified by vertical lines running across the film from the interface to substrates [Fig.~\ref{Fig_S02}\textcolor{blue}{(d)}]; and also, they have lower RT $\mu_e$ in the range of 75–100 cm$^2$ V$^{-1}$ s$^{-1}$~[\textcolor{blue}{1}, \textcolor{blue}{2}].
\begin{figure*}[!h]
     \includegraphics[width=1\textwidth]{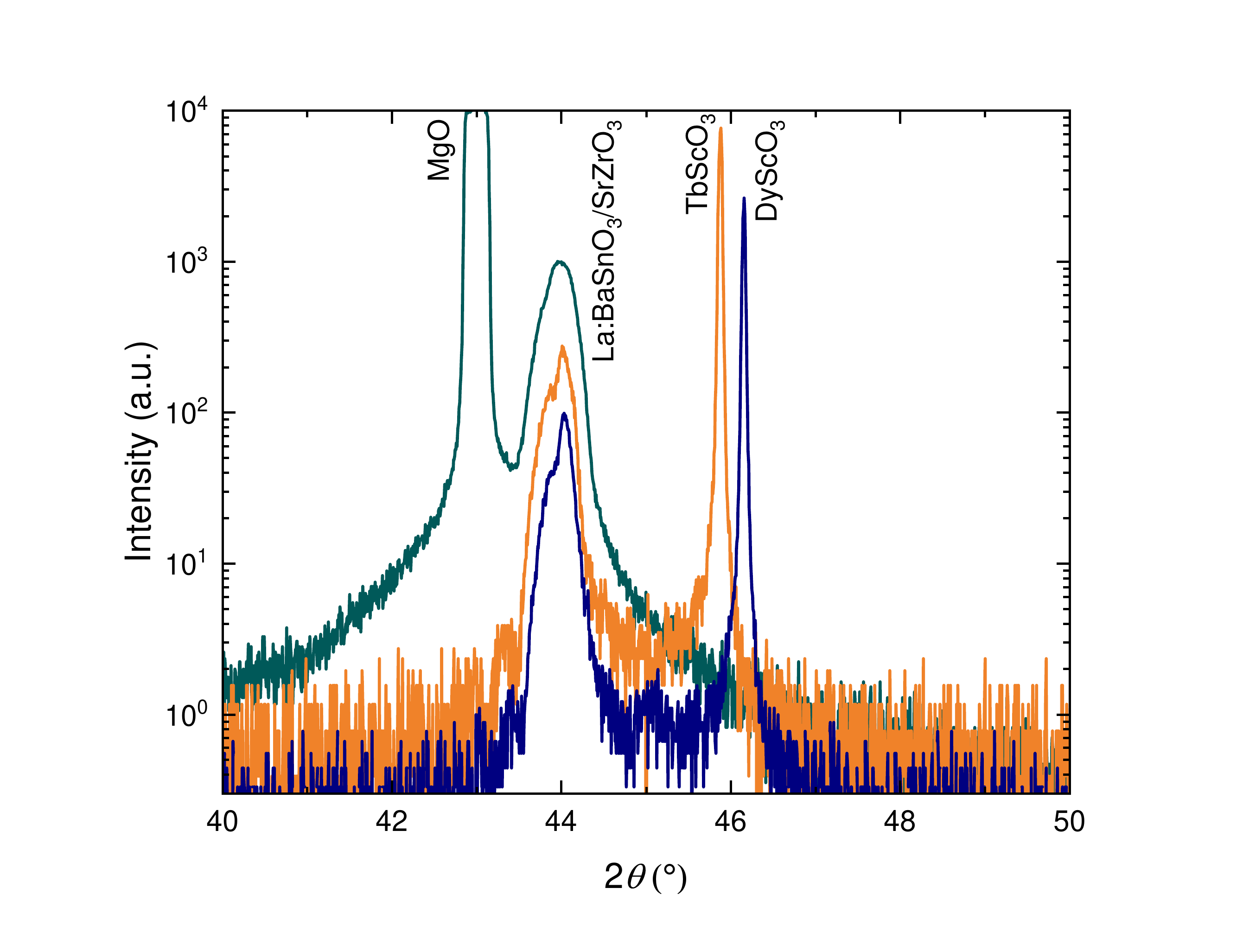}
    \caption{\small{A closeup view of the $\theta-2\theta$ scan around the 002 pseudocubic  peak for the scans of La:BaSnO$_3$/SrZrO$_3$ heterostructures grown on $(110)$-oriented DyScO$_3$ and TbScO$_3$, and $(001)$-oriented MgO  crystalline substrates. The asymmetry in the peaks highlight the presence of the La:BaSnO$_3$ and SrZrO$_3$  layers in these heterostructures. In particular, the 002 peaks of heterostructures grown on  DyScO$_3$ and TbScO$_3$ show noticeable thickness fringes. }} 
\label{Fig_S03}
\end{figure*}

\begin{center}
\textbf{\large{References}}
\end{center}
[1]. A. P. Nono-Tchiomo, W. Braun, B. P. Doyle, W. Sigle, P. van Aken, J. Mannhart, and P. Ngabonziza,\newline \indent \,\,\,\href{https://doi.org/10.1063/1.5094867}{\color{blue}APL Mater. \textbf{7}, 041119 (2019)}{}. \newline
[2]. A. P. Nono Tchiomo, E. Carleschi, A. R. E. Prinsloo, W. Sigle, P. A. van Aken, J. Mannhart, P. Ngabonziza, and \newline \indent \,\,\, B. P. Doyle,  \href{https://doi.org/10.1063/1.5094867}{\color{blue}AIP Advances \textbf{12}, 105019 (2022)}{}.

\end{document}